# *Agito ergo sum*: correlates of spatiotemporal motion characteristics during fMRI

**Authors and affiliations**
Thomas A.W. Bolton[1,2], Daniela Zöller[1,2,3], César Caballero-Gaudes[4], Valeria Kebets[2,5], Enrico Glerean[6], and Dimitri Van De Ville[1,2]

[1] Institute of Bioengineering, Ecole Polytechnique Fédérale de Lausanne, Lausanne, Switzerland
[2] Department of Radiology and Medical Informatics, University of Geneva, Geneva, Switzerland
[3] Developmental Imaging and Psychopathology Laboratory, Office Médico-Pédagogique, Department of Psychiatry, University of Geneva, Geneva, Switzerland
[4] Basque Center of Cognition, Brain and Language, San Sebastian, Spain
[5] Singapore Institute for Neurotechnology, National University of Singapore, Singapore
[6] Department of Computer Science, Aalto University, Helsinki, Finland

**Corresponding author**
Thomas A.W. Bolton
Address: Campus Biotech, Chemin des Mines 9, 1202 Geneva, Switzerland
Phone: +41 21 69 55 254
Email: thomas.bolton@epfl.ch

**Running title**
spatiotemporal motion correlates during fMRI

## Abstract

The impact of in-scanner motion on functional magnetic resonance imaging (fMRI) data has a notorious reputation in the neuroimaging community. State-of-the-art guidelines advise to scrub out excessively corrupted frames as assessed by a composite framewise displacement (FD) score, to regress out models of nuisance variables, and to include average FD as a covariate in group-level analyses.

Here, we studied individual motion time courses at time points typically retained in fMRI analyses. We observed that even in this set of putatively clean time points, motion exhibited a very clear spatiotemporal structure, so that we could distinguish subjects into four groups of movers with varying characteristics.

Then, we showed that this spatiotemporal motion cartography tightly relates to a broad array of anthropometric, behavioral and clinical factors. Convergent results were obtained from two different analytical perspectives: univariate assessment of behavioral differences across mover subgroups unraveled defining markers, while subsequent multivariate analysis broadened the range of involved factors and clarified that multiple motion/behavior modes of covariance overlap in the data.

Our results demonstrate that even the smaller episodes of motion typically retained in fMRI analyses carry structured, behaviorally relevant information. They call for further examinations of possible biases in current regression-based motion correction strategies.

Resting-state functional magnetic resonance imaging (RS fMRI) has been a vibrant and flourishing research topic. Since its advent (Biswal et al. 1995), the assessment of statistical interdependence between brain regions, or *functional connectivity* (FC), has enabled the determination of large-scale functional brain networks (Damoiseaux et al. 2006, Power et al. 2011, Yeo et al. 2011), and the harvesting of their spatiotemporal properties towards a refined understanding of a constellation of brain disorders (Fox and Greicius 2010).

One of the most remarkable features of RS fMRI is that such analyses are already feasible from a few minutes of acquisition (Van Dijk et al. 2009). However, the reliance on low amounts of data also requires that the acquired time courses be impeccably cleaned from potential confounding signals. This is even more of a concern as the field starts moving towards time-varying and -resolved analyses, such as dynamic FC (Laumann et al. 2016; see Preti et al. 2018 for a review) or real-time neurofeedback (Watanabe et al. 2017).

Amongst confounding signal sources, in-scanner head motion of the volunteers has been a leading cause of investigation. Its deleterious impacts may take many forms, and remain incompletely understood (see Power et al. 2015, Caballero-Gaudes and Reynolds 2017 for reviews). Some years ago, it was discovered that even short-lived episodes of motion might greatly bias FC analyses (Power et al. 2012, Van Dijk et al. 2012, Satterthwaite et al. 2012), and lead to erroneous interpretations in clinical or developmental studies (Deen and Pelphrey 2012, Makowski et al. 2019). These observations of motion-biased results further fueled the development of robust post-processing strategies to free fMRI time courses from confounding motion effects.

Thanks to many rigorous and extensive studies (Satterthwaite et al. 2013, Yan et al. 2013, Power et al. 2014, Burgess et al. 2016, Ciric et al. 2017, Parkes et al. 2018), the field has reached a consensus as to what general steps are essential for a viable RS fMRI denoising pipeline. Their specificities, however, remain debated. In short, following the linear realignment of functional images, estimates of motion over time are obtained along three translational directions (left/right, anterior/posterior and dorsal/ventral, respectively termed X, Y and Z in what follows) and three rotational planes (roll, pitch and yaw, respectively referred to hereafter as $\alpha$, $\beta$ and $\gamma$). Framewise displacement (FD) is then computed as an aggregated measure across those 6 motion parameters[1], in order to tag data points corrupted by excessive instantaneous motion and exclude them from subsequent analyses.

Estimated motion time courses are then linearly regressed out from the remaining fMRI data[2], in a matrix of regressors that can be extended to include their quadratic expansions, their derivatives, and/or their squared derivatives.

---

[1] Here, we will be discussing the FD metric suggested by Power et al. (2012), but other alternatives have also been put forward in the past literature (Jenkinson et al. 2002, Van Dijk et al. 2012).

[2] Scrubbed data points can be accounted for in two ways: either by modeling them as individual single-point regressors (Lemieux et al. 2007), or by extracting fitting weights from solely non-scrubbed data points, and then applying the regression to the whole data (Power et al. 2014).

More parsimonious models lead to a greater amount of retained degrees of freedom in the data, while more exhaustive models may remove signal of interest (Bright and Murphy 2015), but enable to account for biophysically relevant nonlinear motion effects (Friston et al. 1996).

Finally, the addition of a covariate for group-level analyses has also been warranted (Ciric et al. 2018). However, this last step has been criticized for its risk of biasing some RS fMRI analyses: indeed, if the behavioral feature of interest in the study positively correlates with the extent of head motion, the investigated metric will be more strongly attenuated in larger movers, thus potentially lowering the true magnitude of the effect of interest.

To date, such concerns have been raised in attention or impulsivity studies (Kong et al. 2014, Wylie et al. 2014). Head motion has been posited to be a marker of cognitive control abilities (Zeng et al. 2014), showing clear heritability (Couvy-Duchesne et al. 2014), even if solely non-scrubbed frames are considered (Engelhardt et al. 2017), and sharing genetic influences with hyperactivity (Couvy-Duchesne et al. 2016) or body mass index (Hodgson et al. 2017). Very recently, an extended multivariate assessment isolated body mass index and weight as the major predictors of head motion, with mild additional impacts of impulsivity levels and alcohol/nicotine consumption (Ekhtiari et al. 2019). Thus, in light of current knowledge, the span of behavioral or clinical measures subject to bias remains limited.

A major limitation of all the above studies, however, is the use of average FD over time to quantify head motion levels. In other words, it is implicitly assumed that motion properties remain similar along the course of a scanning session, and do not differ across translational directions or rotational planes (an assumption that does actually not square well with the meager information available to date; see Wilke 2014). It is likely that the true spatiotemporal complexity of motion is so far overlooked, and that its relationship to behavior is thus only poorly understood. Since even the most sophisticated motion correction approaches summarized above are still unable to fully remove deleterious motion influences (Yan et al. 2013, Siegel et al. 2017), filling such possible gaps of knowledge is a critical question.

Our first question in the present work was thus whether we could find, consistently across subjects, spatiotemporal head motion properties going beyond time- and space-invariance. Our second question was then whether those more subtle motion profiles would consist in endophenotypes characterized by specific anthropometric, behavioral or clinical properties.

## Materials and Methods

**Motion Data Acquisition and Preprocessing**

We considered a set of 224 healthy subjects from the *Human Connectome Project* (Smith et al. 2013), scanned at rest (eyes open) over four separate 15-minute sessions at a TR of 0.72 s. For each session, motion was estimated using rigid-body transformation with three translation parameters (along the X, Y and Z axes) and three rotation angles (in the α, β and γ planes respectively highlighting roll, pitch and yaw) with respect to a single-band reference image (*SBRef*) acquired at the start of each session, and FSL's FLIRT (Jenkinson et al. 2002). It resulted in 6 time courses (one per motion parameter) with 1200 time points each.

In the present work, we solely analyzed the motion time courses (not the fMRI data) from the first acquired session. Individual motion time courses were differentiated so that our analyses would focus on instantaneous displacement from time *t* to time *t+1*. Further, since time points linked to excessive displacement are typically removed from RS fMRI analyses, we only considered non-scrubbed motion instances according to Power's FD definition (Power et al. 2012) at a threshold of 0.3mm. Resorting to a more conservative (0.2 mm) or more lenient (up to 1 mm) threshold, or censoring not only tagged time points (time *t*) but also the following ones (time *t+1*), did not modify our findings (see **Supplementary Material** for a more detailed description).

**Spatiotemporal Motion Characterization**

We wished to assess whether different subjects would present distinct spatiotemporal motion characteristics in the data points that are typically conserved in RS fMRI analysis (*i.e.*, not scrubbed out).

For each motion time course, we computed absolute valued instantaneous displacement. Thus, we did not consider the sign of the changes (*e.g.*, moving positively as opposed to negatively in the X direction); this is because initial analyses indicated that positive-valued and negative-valued movements always compensated, to the exception of the X case (two-sided Wilcoxon rank sum test, p=0.0001).

Then, we averaged motion values within a) each motion type (X, Y, Z, $\alpha$, $\beta$ and $\gamma$) and b) each of 6 even-duration time intervals along the scanning sessions (2.4 min = 144 s each). This resulted in a total of 36 conditions. We chose 6 temporal sub-bins to give equal weight to spatial and temporal domain information in our decomposition of the data. Eventually, the values were z-scored across subjects for each condition so that positive values highlight strong movers (at a given time and for a given motion parameter) with respect to the mean, and *vice versa*. It also follows that an equal weight is given to each condition.

Next, we used these 36 motion summary measures to separate subjects into different subgroups of movers through consensus clustering, a nonlinear dimensionality reduction approach (Von Luxburg 2007; see **Supplementary**

**Material** for details). By taking into account such precise motion characteristics, we exploit complex motion profiles rather than simply dividing into high- and low-motion subjects, as is classically done on the basis of average FD.

To evaluate whether there was any significant effect of scanning duration, motion parameter or mover subtype, or any interaction between these factors, we conducted a three-way ANOVA (factor 1: scanning duration [time], factor 2: motion parameter [space], factor 3: mover subtype [group]) and assessed significance by comparing the obtained F-values with a null distribution generated non-parametrically over 10'000 folds, shuffling the three factors independently from each other.

To assess motion changes along time within a given group of subjects, a linear model was fitted along the six time bins (including a constant regressor of no interest), and the null hypothesis that the mean $\beta$ value across subjects would be equal to 0 was assessed. To examine differences in motion along space, we conducted pair-wise two-tailed t-tests between all group pairs.

**Behavioral Data Acquisition and Processing**

For each subject, a battery of behavioral and demographic scores was also quantified. A list of all the investigated scores in the present study can be found in **Supplementary Table 1**. They were subdivided into several key sub-domains, largely following the original classification found in the HCP Data Dictionary[3]:

1. Demographic parameters, including race, ethnicity, employment status, income or education level
2. Physical health, such as weight, height, body mass index (weight/height$^2$), blood pressure, hormonal levels
3. Alertness levels, assessed in terms of cognitive status (MMSE; Folstein et al. 1983) and sleep quality (PSQI; Buysse et al. 1989)
4. Cognitive abilities (in terms of accuracy, response time or errors) across various tasks spanning different cognitive domains (see Barch et al. 2013 for details)
5. Emotional level in terms of anger, fear, stress or life satisfaction (assessed through the NIH toolbox; Gershon et al. 2010)
6. Motor abilities, including endurance, gait speed, dexterity and strength measurements
7. Sensory levels, quantified in terms of responses to noise, odor, pain, taste, or contrast
8. Personality traits, as assessed by the NEOFAC questionnaire (McCrae et al. 2004)
9. Psychiatric and life function, including for example measures of anxiety, aggressiveness, withdrawal or inattention (Achenbach 2009)
10. Substance use, that is, intake of alcohol, tobacco or drugs (partly from the SSAGA questionnaire; Buchholz et al. 1994).

---

[3] https://wiki.humanconnectome.org/display/PublicData/HCP+Data+Dictionary+Public-+Updated+for+the+1200+Subject+Release

For some scores, several entries were not acquired in a sub-fraction of subjects (mean: 1.52%, median: 0.89%, maximum: 21.43%). This was taken into account in behavioral data processing so that it would exert a minimal effect on the described findings. Some scores were also discarded due to various criteria, and the remaining ones were processed as in Smith et al. (2015), yielding a total of 45 scores for subsequent analyses, reflective of anthropometric properties, cognitive abilities or clinical features. Details are provided in the **Supplementary Material**.

**Univariate link between motion subgroups and anthropometry/behavior**
To determine whether some anthropometric/behavioral scores would differ across mover subgroups, we performed a univariate assessment. For each of the 45 assessed domains, we computed a score indicative of cluster-to-cluster distinction. Formally, following Gu et al. (2012):

$$F(x^i) = \frac{\sum_{k=1}^{K} n_k (\mu_k^i - \mu^i)^2}{\sum_{k=1}^{K} n_k (\sigma_k^i)^2},$$

where $x_i$ is the vector of the $i^{th}$ domain scores across subjects, $\mu^i$ is its average regardless of group classification, $\mu^i_k$ is its average within group k, and $\sigma^i_k$ is the standard deviation within group k. A large score value indicates that the assessed behavioral domain shows distinct values between clusters.

To non-parametrically extract significant scores, we used permutation testing, by randomly shuffling subject motion entries 1'000 times. P-values were Bonferroni corrected for 45 tests. Scores were considered significant at a corrected p-value of 0.05.

**Multivariate Links Between Motion Features and Anthropometry/Behavior**
To go beyond univariate comparisons and test for multivariate patterns of motion-behavior interactions, we conducted a Partial Least Squares (PLS) analysis (McIntosh and Lobaugh 2004, Krishnan et al. 2011). We summarize the gist of the approach below, and additional details can be found in the **Supplementary Material**.

We considered the matrix of behavioral scores (size 224 x 46) on the one hand, and the matrix of spatiotemporal motion features (size 224 x 36) on the other. Using PLS, we derived a set of so-called *components*. Each consists in a linear combination of motion scores, and a linear combination of behavioral scores, with maximized covariance. The associated weights are termed *motion saliences* and *behavioral saliences*, and are respectively arranged in **U** and **V**, two matrices of size 36 x 36 and 46 x 36. Motion saliences (*i.e.*, the columns of **U**) are orthonormal, and so are behavioral saliences. Successive components explain gradually less of the covariance present in the data, as quantified by their *singular values*. Finally, the extent to which a motion salience or a behavioral salience is expressed in a given subject is termed the *motion latent weight* or *behavioral latent weight*, respectively.

To assess significance of the PLS components, we compared their singular values to a null distribution constructed from 1'000 shuffled datasets, following Zöller et al. (2017). We focused our interpretation on the components significant at p=0.02. To determine the stability of the saliences, we performed bootstrapping with 80% of the data.

For interpretation, we converted the 36-element motion saliences obtained from PLS analysis into a 6-element space and a 6-element time representation, by averaging across all time points or across all spatial directions, respectively. Stability was assessed on those summarizing values. For each behavioral or motion salience element, we considered it significant above a bootstrap score (mean salience across bootstrapping folds divided by the associated standard deviation) of 3, corresponding to a confidence interval of approximately 99% (Garrett et al. 2010; Zöller et al. 2017).

In addition, we performed correlation analyses between motion (or behavioral) latent weights of the analyzed components and FD or age, using Spearman's correlation and non-parametric significance assessment. We also performed a Wilcoxon rank sum test to probe for possible differences in motion (or behavioral) latent weights across gender. Results were Bonferroni corrected for 18 tests (3 components examined in terms of 3 separate parameters for 2 types of latent weights) and judged significant at a corrected p-value of 0.05.

## Results

**Spatiotemporal Motion Diversity**

Average motion across six even-duration time bins, and the 6 motion parameters, was quantified. This spatiotemporal motion profile characterization revealed the existence of four separate subgroups of movers (**Figure 1A**): in the first one ($n_1$=70, red patches), subjects showed low motion across all time and motion dimensions (negative z-score values in **Figure 1C**). In the second ($n_2$=51, dark blue patches), subjects moved little, and less following the first sixth of the session, with stronger motion along the $\alpha$ and $\beta$ rotational components. The third group ($n_3$ = 67, orange patches) showed very strong motion spatiotemporally, which increased after the first sixth of the session, while the fourth one ($n_4$ = 36, cyan patches) showed particularly strong motion in the $\gamma$ rotational plane, which slightly attenuated after the first sixth of the session.

In addition, we also individually plotted scanning duration or motion parameter against cluster assignments (**Figure 1C**), averaging over all entries from the other factor (*e.g.*, the bar labeled 'X' denotes the average of motion along the X direction from $t_1$ to $t_6$).

Statistical analysis confirmed the above observations: on top of a significant effect of group (F = 1414.41, $p<10^{-5}$), there was a significant time x group interaction (F = 3.11, $p<10^{-5}$), and *post-hoc* assessment revealed that while groups 1, 2 and 4 showed a decrease in motion over time ($\beta_1$=-0.007 [-0.01,-0.003], p=0.0026; $\beta_2$=-0.0194 [-0.027,-0.012], $p<10^{-5}$; $\beta_4$=-0.025 [-0.031,-0.019], $p<10^{-9}$), group 3 exhibited an increase ($\beta_3$=0.0354 [0.014,0.057], p=0.0013). Thus, different mover subgroups displayed varying temporal changes in their extent of motion.

In terms of spatial properties, there was a significant effect of space (F=5.92, $p<10^{-5}$), as well as a significant space x group interaction (F = 83.88, $p<10^{-5}$). Exhaustive results from a *post-hoc* assessment are displayed in **Supplementary Table 2**. They show that subjects in group 4 moved the most in the γ plane (hence their blue shade in **Figure 1B**), while subjects from group 2 moved the least on that plane (hence their red and green tones). Group 1 featured the lowest movers in Y, Z, $\alpha$ and $\beta$, while in group 3, subjects moved most in X, Y, Z, $\alpha$ and $\beta$ (thus, they appear in white in **Figure 1B**). Overall, each group could thus be clearly distinguished on the basis of spatial motion properties.

**Univariate Links Between Motion and Anthropometry**

Next, we related the spatiotemporal motion characteristics of the subjects (as summarized by their mover group assignment) to their anthropometric, behavioral and clinical features. **Height**, **Weight**, **Blood pressure** and **Cognitive flexibility** scores were significantly different across mover subtypes following Bonferroni correction (**Figure 2**). When applying FDR correction instead, **Endurance**, **Somatic problems** and **Inattention** also became significant.

Subsequent inspection of pair-wise group relationships (**Figure 2, inset**) showed that group 3 (*i.e.*, the largest movers) showed greater weight, lower height, more elevated blood pressure and reduced endurance compared to all other groups, highlighting that they largely stand out in terms of anthropometric features. Members from group 4 (that is, the γ movers) mostly differed from groups 1 and 2 through larger somatic problems and inattention scores; hence, they can rather be distinguished on the basis of clinical measures (note that no pair-wise comparison survived Bonferroni correction in this case; this is unsurprising given that clinical scores were only found significant upon FDR correction in the above evaluations). Finally, subjects from group 1 (the lowest movers) showed significantly larger cognitive flexibility compared to groups 2 and 3, hence differing in terms of behavioral abilities.

**Subtler Motion/Behavior Relationships Revealed by Multivariate Analysis**
Finally, we attempted to extract significant multivariate relationship(s) between our spatiotemporal motion characteristics and the entire breadth of anthropomorphic/behavioral/clinical features (**Figure 3**).

There were three significant covariance components. Component 1 ($p<0.001$) explained 71.92% of the data covariance, and characterized motion regardless of space or time (apart from $\alpha$, all values had an absolute bootstrap score larger than 3): subjects showing larger global motion values also showed worse endurance, larger sensitivity to auditory noise, worse working memory performance, worse spatial orientation (both in terms of larger response time and lower accuracy), lower cognitive flexibility, worse fluid intelligence, and worse body mass index (decreased height, enhanced weight). In addition, they were more anxious, showed more thought problems, aggressiveness, inattention and antisocial behaviors. When compared to the mover groups derived beforehand, the gradient in motion latent weights across subjects appeared to discriminate low movers (group 1) from large movers (group 3).

Component 2 ($p=0.017$) explained 15.2% of the covariance of the data. It largely corresponded to movement in the $\gamma$ plane independent from time (none of the t-values were significant). Stronger movers in that plane also showed sleep problems, but performed better in cognitive flexibility and spatial orientation tasks. Their response time was larger when tested for sustained attention, and they showed a constellation of elevated clinical scores including anxiety, withdrawal, somatic problems, attentional problems, aggressiveness, hyper-responsiveness and rule breaking behaviors. They also showed stronger alcohol consumption. Motion latent weights contrasted the γ movers (group 4) from other groups.

Component 3 ($p=0.008$) explained 5.35% of data covariance, and contrasted translational (mostly X) and rotational (mostly $\alpha$) motion. Stronger translational and lower rotational movers showed greater weight and height, higher blood pressure, enhanced cognitive flexibility abilities, greater self-regulation abilities, and larger response time upon a sustained attention task. They were also better at discriminating emotions, but worse at contrast sensitivity. Clinically, they had more intrusive thoughts and showed less externalizing, withdrawal and thought problems. As for motion latent weights, they separated the α/β movers (group 2) from the others.

Motion latent weights of component 1 positively correlated with FD (**Figure 4**; R=0.87, p=0), and so did behavioral latent weights (R=0.49, p=0). For component 2, there was a negative correlation between motion latent weights and FD (R=-0.31, p=0), as well as negative correlations between motion/behavioral latent weights and age (respectively, R=-0.23, p=0.018/R=-0.21, p=0.022). Finally, motion and behavioral latent weights of component 3 showed a significant gender difference (male > female, t=4.01, t=8.83, respectively).

## Discussion

**A cartography of in-scanner spatiotemporal motion**

In RS fMRI analyses, time points associated with excessive instantaneous motion (as quantified by a composite FD score) are typically removed/scrubbed. Two implied assumptions are made through this process: the first is that motion in the retained, non-scrubbed time points is negligible and does not show any characteristic structure. The second assumption is that FD is sufficient to characterize motion.

Our results question both of the above assumptions. Indeed, we were able to separate scanned volunteers into four different mover subgroups even when excluding time points with high motion as quantified by FD. These groups differed in the extent of motion displayed by subjects along the three translational (X, Y and Z) and three rotational ($\alpha$, $\beta$ and $\gamma$) motion directions, as well as in the temporal evolution of motion along the scanning session.

More specifically, a change in motion extent along the scanning session was observed, as seen by a significant time x group interaction in our ANOVA design, as well as by significantly non-null *post-hoc* regression coefficients. The absence of time x space, or time x space x group interaction terms means that temporal changes were consistent across all motion directions. Further, a closer inspection revealed that the main drive of this result was a changed motion extent after the first sixth of the session (see **Figure 1C**): in subjects from group 3, motion increased after the first 2.4 min of recordings, while the opposite was seen for the other groups. This need for a few minutes before setting into a *motion steady state* suggests that to avoid one possible source of bias, if affordable, future fMRI analyses may ignore the first few minutes of a scan.

In addition to temporal motion properties, there were also much more noticeable group differences in terms of space, which contributed in major part to the organization revealed in the dimensionally reduced representation of **Figure 1A/B**. Two mover subgroups were opposite extremes: group 1 subjects moved very little as compared to the average across spatial directions (hence depicted in black in **Figure 1B**), while the subjects in group 3 consistently displayed very large displacements (hence represented in white). Rotational motion was the distinguishing feature of the last two groups, respectively in the $\alpha/\beta$ (group 2, seen as red and green shades) and the $\gamma$ (group 4, color coded by a blue shade) planes.

All in all, the presence of strongly differing spatial motion profiles across subjects confirms the importance of subject-level motion correction strategies through regression. Further, since our work focused on instantaneous motion (*t* to *t+1* changes), our results may be interpreted as an additional argument in favor of more complete regression models, at least to the point of incorporating motion time courses and their shifted counterparts (see below, however, for a more complete discussion).

Rather than considering four subgroups of movers, another perhaps more relevant interpretation of our data is the presence of three distinct axes of motion: the first is a global component, homogeneous across space. It is seen as the first dimension in our spectral clustering investigations (**Figure 1A/B)**, positioning it as the major discriminating factor across subjects. It is also found back as the first, most prominent component in the following PLS analysis (**Figure 3A**, left panel). Other more subtle, but nonetheless significant spatial motion combinations then add up on top: motion along the $\gamma$ plane and in the $\alpha/\beta$ planes are, respectively, captured in dimensions 2 and 3 in spectral clustering, and in components 2 and 3 in the PLS analysis.

The two methodological approaches employed strongly differ in their core properties: spectral clustering is a hard clustering technique in which each subject can only be assigned to one group, while PLS describes the motion properties of each subject as a linear combination of motion saliences that co-vary with anthropometry, behavior and clinical symptoms. The convergent findings obtained from both analytical perspectives strengthen our newly revealed cartography of spatiotemporal movers as a non-negligible feature.

Our findings question the accuracy of most motion correction assessment approaches, which exclusively rely on averaged FD over time. A separate assessment across motion parameters (or more elaborate approaches involving specific weighted combinations, such as our PLS motion saliences) appears to be necessary to better understand which motion impacts are removed, and which subsist in the data.

***Agito ergo sum*: bodily and behavioral underpinnings of motion**

In 1644, René Descartes, in quest for a primal principle at the root of all knowledge, formulated his notorious *cogito ergo sum* (I think, therefore I am)[4]. 375 years later, we wish to summarize our findings by reformulating his words: ***agito ergo sum*** (I move, hence I am). By this, we mean that the defining aspects of someone (one's bodily features, abilities to interact with the world and ways to respond to the environment around) are reflected, in various and subtle ways, in how one moves during scanning.

As an example of this principle, while subjects from groups 1, 2 and 4 moved less after the first sixth of the recording session, high movers from group 3 moved more. Univariate evaluation following spectral clustering revealed that the latter mostly stood out in terms of anthropometric or fitness measures (**Figure 2**, inset): weight, height, blood pressure and endurance. Conversely, group 1 subjects stood out by their better cognitive flexibility scores.

Those observations enable to sketch a global picture of how the response to the scanning environment differs across subjects: low movers, who are able to efficiently cope with changes in environmental conditions (for example, by better adjusting to the loud MRI noise fluctuations), rapidly start moving less and maintain overall low head motion throughout scanning. Large movers, on the other hand, are intrinsically more prone to large head motions, possibly because of feeling more cramped inside the scanner, and become increasingly uneasy with the contiguous MRI environment, thus moving more.

A caveat of univariate approaches is the risk that more subtle behavioral or clinical correlates of motion remain undetected. Our follow-up multivariate PLS analysis confirmed this limitation: motion saliences from the most prominent component (**Figure 3A**, left panel) were positive across space and time, with an increase following the first session sixth. This means that subjects expressing this component more strongly move more overall, and *vice versa*, as also confirmed by a strong positive correlation with FD (**Figure 4A**). This component thus highlights similar motion features as the ones discriminating mover groups 1 and 3. The array of associated behavioral saliences not only included the dominating anthropometric factors mentioned above, but also showed that larger movers perform worse in working memory, fluid intelligence and spatial orientation scores. Further, they also exhibit worse aggressiveness, inattention and antisocial behavior clinical scores.

Overall, this global pattern is highly reminiscent of a positive-negative mode of population covariation previously described by Smith and colleagues (2015), and put forward as relating behavior, demographics and FC. The similarity may partly come from the fact that the authors resorted to Canonical Correlation Analysis (CCA), a multivariate technique with strong similarities to PLS. Our results raise the possibility that this mode may, at least in part, reflect differences in motion across the considered subjects.

---

[4] The first mention of that particular formulation indeed dates back from the *Principia philosophiae,* published in 1644.

The main PLS component highlights the dominating factor of motion/behavior covariance. On top of it, we also revealed subtler overlapping factors. Component 2 (**Figure 3B**) specifically showcased the $\gamma$ motion seen in group 4: those subjects that express it strongly move a lot along $\gamma$, but also move less along other directions (as indicated by the negative signs in **Figure 3B**, left panel). This is why motion saliences were negative (albeit non significantly) across time bins, and why a negative correlation was found between motion latent weights and FD (**Figure 4B**). Component 2 may thus be seen as a positive marker of low motion.

No significant anthropometric associations were detected, but the subjects expressing component 2 more strongly showed sleep disturbances, good cognitive flexibility and spatial orientation performances, as well as greater carefulness in an attention response task (better sensitivity at the cost of larger response times; see **Supplementary Material**). This was accompanied by a wide scope of elevated clinical scores, including anxiety, somatic problems, aggressiveness, intrusiveness, ADHD and hyper-responsiveness, as well as by marked alcohol consumption.

We conjecture that this component may reflect efforts of the subjects to refrain from moving in the scanner: indeed, head motion along γ reflects yaw, and may highlight attempts at limiting translational displacements along X or Z by forcing the head to remain anchored on the bed. The efforts leading to this typical motion signature may be regulated by the subjects' good cognitive abilities, and were perhaps influenced by their clinical symptoms.

Interestingly, the expression of component 2 also negatively correlated with age, despite considering a relatively narrow age range in the present study (between 26 and 35 years old). Thus, older subjects express this component less, and move more. Since head motion has been a central question in developmental studies, it will be interesting to examine, in future work, whether the characterization of motion along γ rather than through FD may be a better strategy (especially given that component 1, accounting for the global motion effect, showed no significant relationship to age).

As for component 3 (**Figure 3B**), it contrasted motion along X and α (*i.e.*, more strongly expressing subjects move more along X, but less along α). Positive motion and behavioral latent weights were seen in males, while the opposite was seen for female subjects (**Figure 4C**), implying that gender may be an underlying cause of that particular motion pattern. α reflects roll, occurring in the plane spanned by the X axis: motions along α and X are thus biophysically constrained to occur concurrently. The differential recruitment of both motions across genders may result from distinct anthropometric factors (larger weight, height and blood pressure in males), or from behavioral/clinical specificities of one of the genders (for example, better self-regulation abilities in males).

**Implications, limitations and future perspectives**

Our results have strong implications regarding RS fMRI studies: indeed, the observation that a broad array of behavioral and clinical characteristics relate to motion implies that the scope of studies reporting possibly biased findings with regard to clinical or cognitive group-level comparisons is perhaps much wider than envisaged so far. On top of previously questioned results regarding fluid intelligence (Finn et al. 2015; see **Figure 6** of Siegel et al. 2017), former reports focusing on sustained attention (Rosenberg et al. 2016) or extraversion (Hsu et al. 2018) may also need to be reconsidered.

Earlier on, we discussed how the widely used extended subject-level regression designs enable to remove the spatiotemporally complex motion effects introduced here. However, their intricate and overlapping relationships with behavior raise the danger that, akin to including average FD as a covariate in group-level analyses, an unwanted bias with regard to clinical or cognitive analyses occurs at the single-subject level stage.

Assume, for example, that an experimenter is interested in studying sleep quality through assessments of FC at rest. From our results, subjects showing worse sleep quality will exhibit greater motion along the $\gamma$ axis, in a way that is essentially constant at the time scale of minutes of recording (**Figure 3B**). The use of a regressor encoding instantaneous motion changes along $\gamma$, as suggested by most for optimal data preprocessing, may result in the removal of a larger signal fraction in individuals with poor sleep quality, possibly leading to the underestimation of the effect of interest. For this reason, we encourage experimenters, in future analyses, to investigate the fitting coefficients obtained upon regression so that it can be verified whether a link exists between the extent of removed signal, and the behavioral feature of interest.

Of course, the exact impact of the regression step will depend on the precise temporal expression of $\gamma$ motion and of sleep-related fMRI fluctuations, since one fitting coefficient is extracted depending on frame-wise similarities between the considered motion and the voxel-wise fMRI time courses. The obvious next step to perform, and the major limitation of the present analyses, is that we have not yet pushed our exploration to the level of fMRI time courses, but focused on motion estimates only.

Our aim, with this report, was not to design a new efficient motion correction strategy, but to dig into the complexity of motion *per se*, and by this mean, put forward possible caveats and improvements of existing approaches. Our code and results are fully available at https://c4science.ch/source/MOT_ANA.git, and we encourage the interested researchers to extend our current investigations at the level of the fMRI signal.

A second limitation of our work is that we solely considered motion, although many more factors are known to corrupt the fMRI signal (Biancardi et al. 2009, Birn 2012, Liu 2016). Particularly relevant to the present study is the recent work of Power et al. (2019), who showed that motion time courses from the HCP dataset contain an array of respiratory contributions. Given the impact of blood pressure on some of our components, it seems likely that cardiac or respiratory effects indeed contribute to head motion variability.

It is important to specify that although regression-based approaches are one of the major preprocessing avenues, other motion correction alternatives also exist and may less suffer from possible biases; they include original twists on traditional regression designs (Patriat et al. 2015, Patriat et al. 2017), more sophisticated variants over scrubbing (Patel et al. 2015, Yang et al. 2019), and methods relying on an ICA decomposition of the data (Salhimi-Korshidi et al. 2014, Pruim et al. 2015).

Future motion correction strategies shall improve over current ones in several ways: first, through more elaborate acquisition schemes, such as with multi-echo sequences (Power et al. 2018); second, through the exploration of other complementary denoising strategies, such as with fMRI simulators (Drobjnak et al. 2006) or prospective correction (Zaitsev et al. 2017); third, and perhaps most importantly, through an efficient cross-talk across those strategies. For example, it was recently shown that the use of customized head molds reduces motion during scanning on young subjects (Power et al. 2019); this could be pushed further by orienting the design in subject-specific manner, using motion characteristics such as the ones described here.

## Conclusion

We demonstrated that head motion in the MR scanner during RS fMRI acquisitions, an infamous confounding factor of this imaging modality, exhibits spatiotemporal structure that is not fully accounted for by motion-correction strategies. Strikingly, one's motion characteristics can inform not only about one's anthropometry, but, more surprisingly, about one's behavior and psychiatric function. We hope that our findings will lead future clinical or cognitive fMRI studies to probe more extensively for the presence of motion-related artifacts.

## Acknowledgments

The authors would like to thank Stefano Moia for his help on interpreting individual motion parameters, and Raphael Liégeois for his rereading of the manuscript and insightful suggestions.

## Authors' contributions

TB designed the study, ran the analyses and wrote a first draft of the manuscript. DZ provided the PLS software used in the work, as well as methodological insights. CC and EG suggested extra analytical steps. VK contributed to the cognitive interpretation of the findings. DVDV supervised the work. All authors helped in writing the final manuscript version.

**Figure legends**

**Figure 1: groups of spatiotemporal movers. (A) (Top)** Proportion of ambiguously clustered pairs (PAC) across different evaluated numbers of clusters. The color gradient from black to yellow denotes PAC evaluation for an increasingly narrow distribution range. Lower values highlight stronger robustness of clustering, and the optimum (K=4) is labeled by an arrow. **(Bottom)** Dimensionally reduced representation of all 224 subjects, each depicted by a three-dimensional box. Box width along the first, second and third dimension are proportional to the average motion extent, across all 6 considered time bins, in the X, Y and Z directions. Colors denote the four different subgroups of movers. Edge thickness of the boxes is proportional to the slope of a linear fit to average spatial motion over the 6 temporal bins, while red/blue symbolize increased/decreased motion over time. **(B)** Similar representation, with color coding in RGB scale proportional to the extent of motion in the $\alpha$ (red), $\beta$ (green) and $\gamma$ (blue) rotational planes. Black/white denotes uniformly low/high motion along the three rotational planes. **(C)** Simplified representation of the data along time and clusters (top row), or along space and clusters (bottom row).

**Figure 2: univariate links between spatiotemporal motion and anthropometry/behavior/clinical scores.** Across all 45 considered domains, Fisher score in terms of discriminability across the four spatiotemporal mover groups. Black horizontal bars denote significance thresholds, derived non-parametrically and Bonferroni-corrected for 45 tests. Grey bars denote significance thresholds upon FDR correction. **(Inset)** For the 7 significant domain scores, *post-hoc* comparison of Fisher score values across mover subgroups. Positive values highlight stronger scores for the row group. One star (*) highlights significance without multiple testing corrections, and two stars (**) highlight significance upon Bonferroni correction for 42 tests (6 pair-wise comparisons for 7 scores).

**Figure 3: motion/behavior covariance components.** Z-score reflecting the motion (left) and behavioral (middle) saliences of the first **(A)**, second **(B)** and third **(C)** components extracted from a PLS analysis. Dashed horizontal lines denote the significance threshold (|bootstrap score|>3), and text labels are appended to the domain scores showing significance. $t_1$ to $t_6$ represent the first to sixth temporal bins of a session. For each component, motion latent weights are also color-coded on two views from a dimensionally reduced representation of spatiotemporal motion as in **Figure 1** (right).

**Figure 4: relationship of latent weights with FD, age and gender.** For the first **(A)**, second **(B)** and third **(C)** components, significant correlations of motion or behavioral latent weights with FD or age, and significant differences across gender. Provided p-values are Bonferroni-corrected for 18 tests.

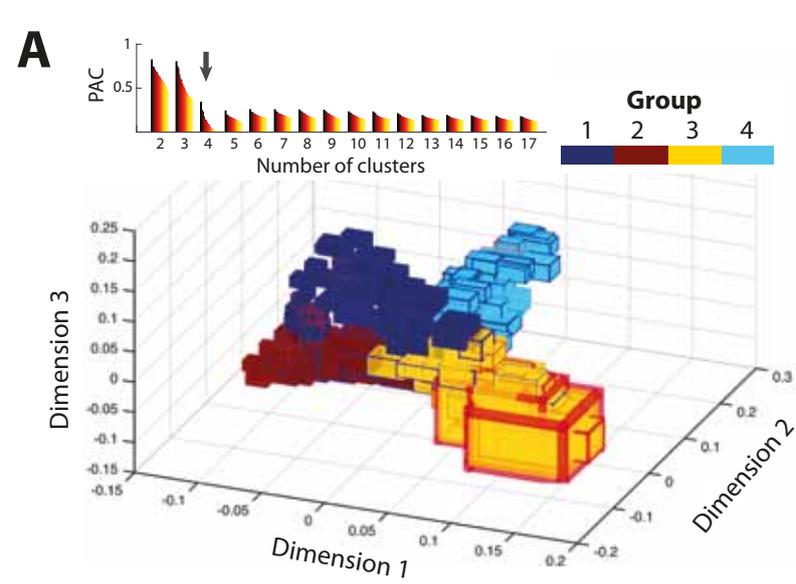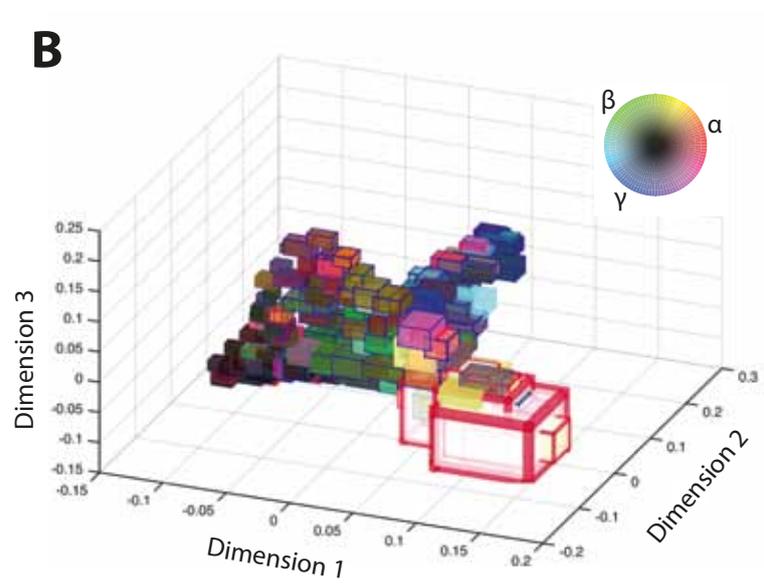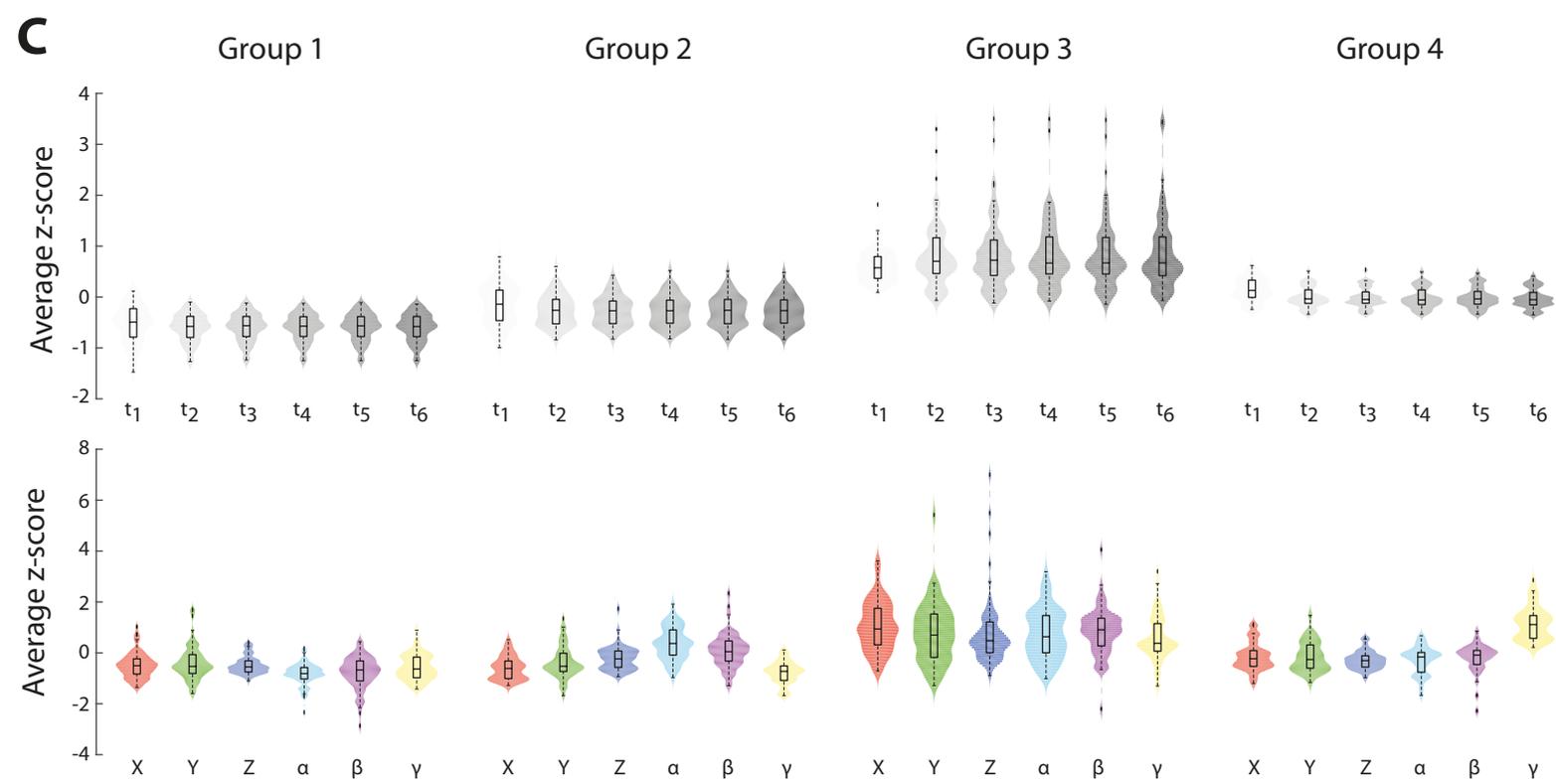

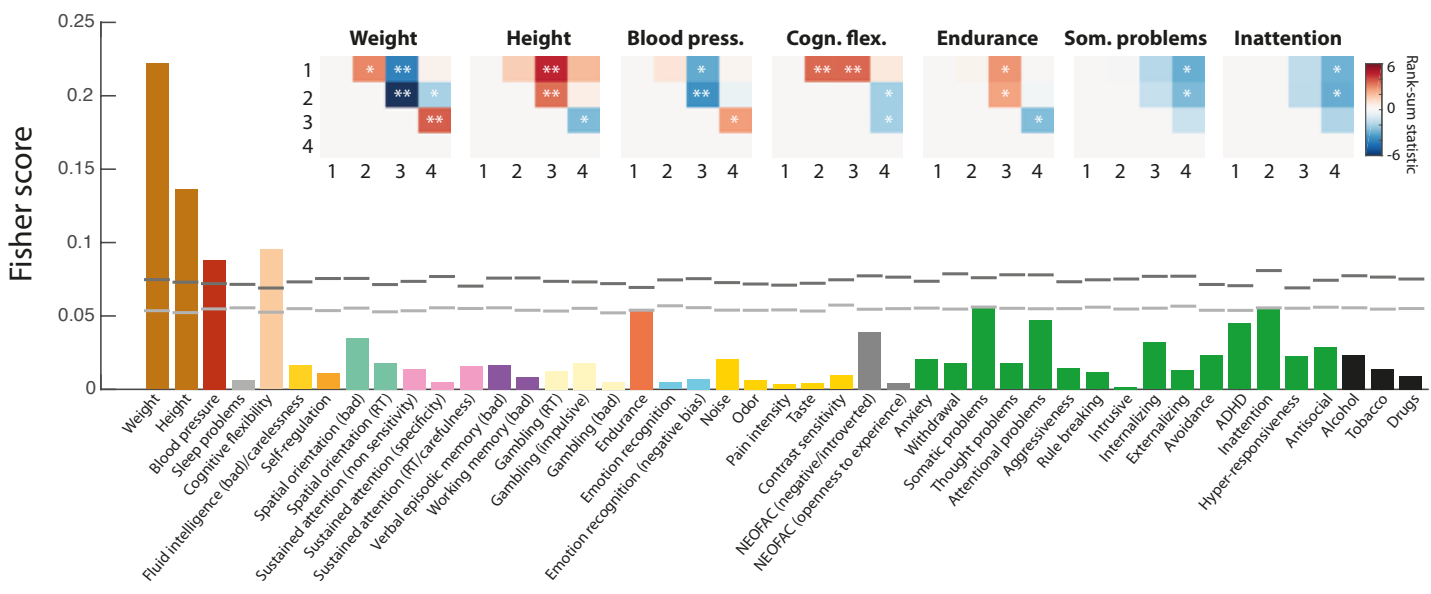

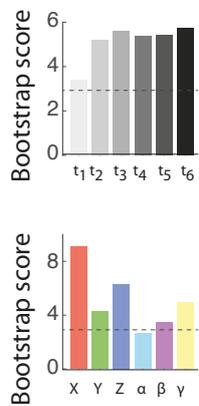
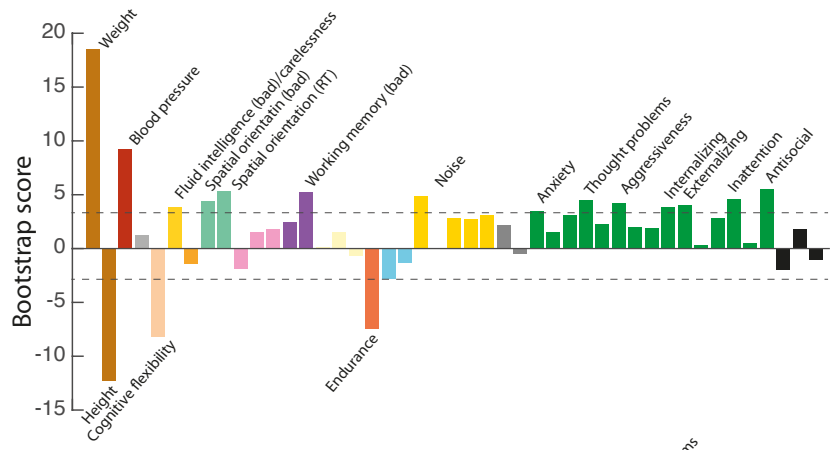
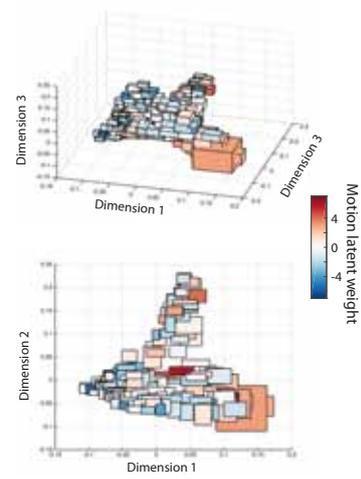
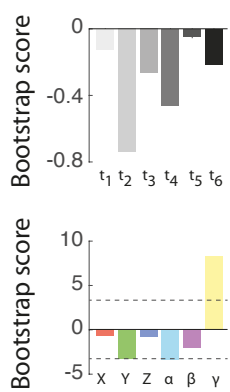
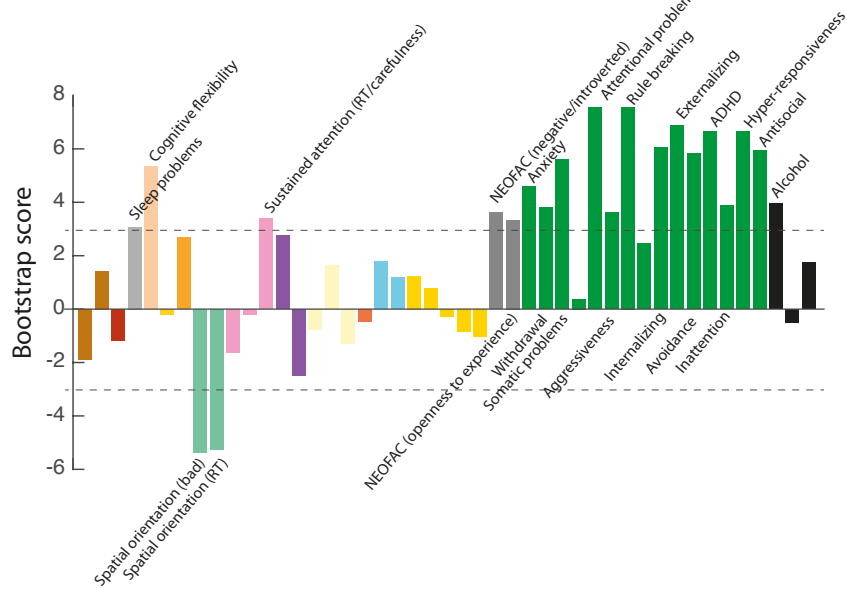
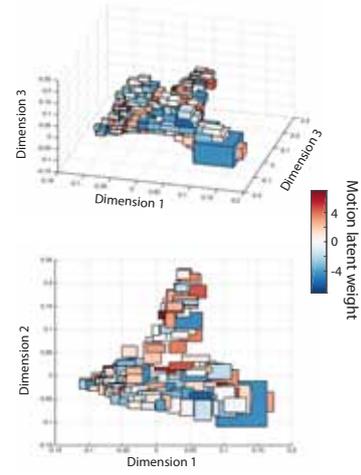
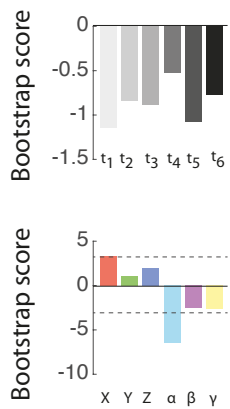
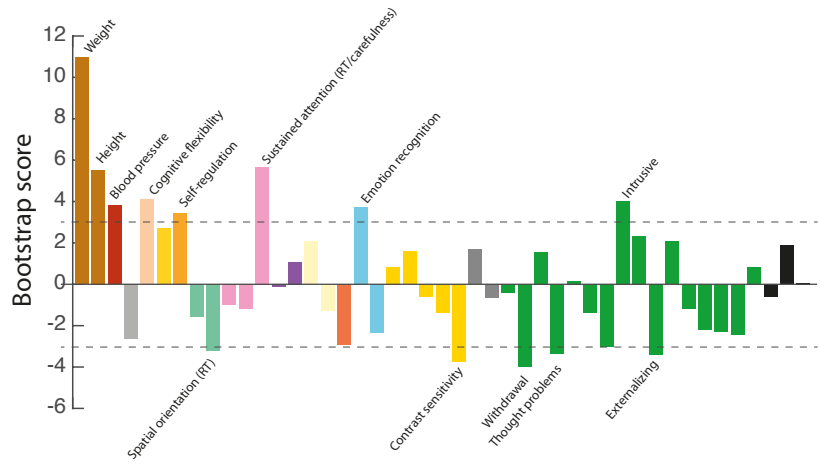
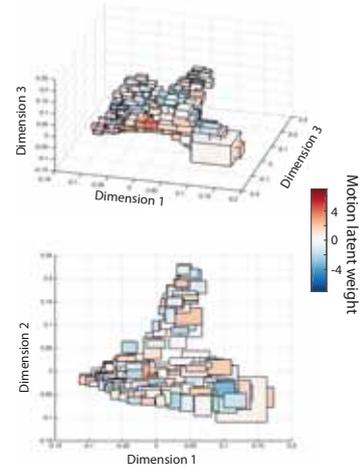

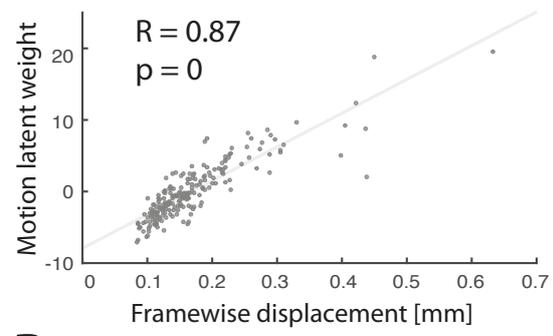
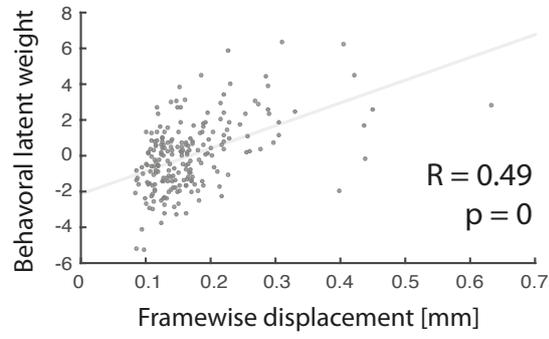
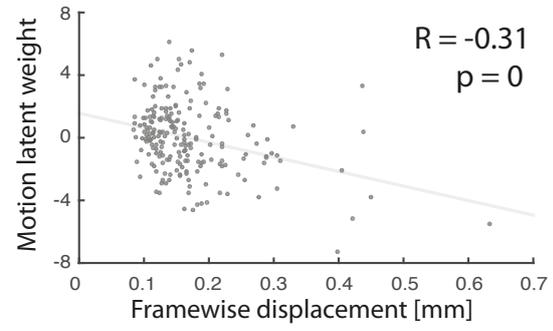
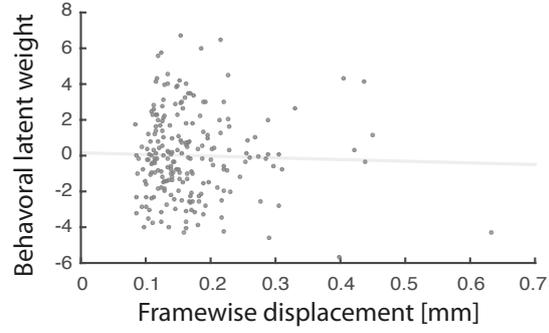
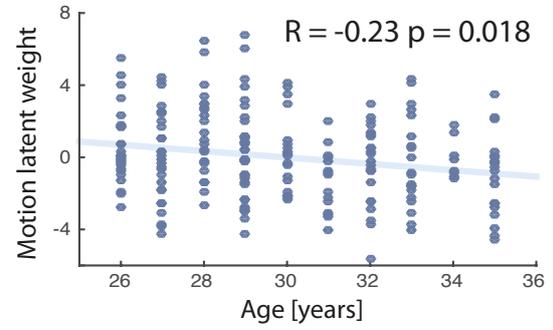
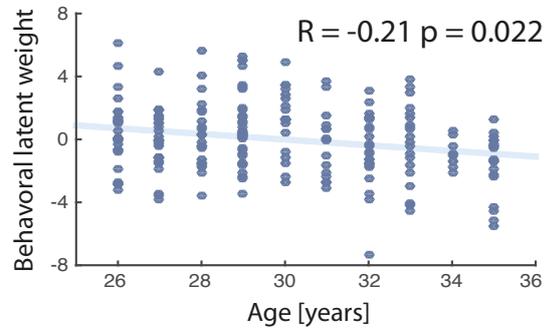
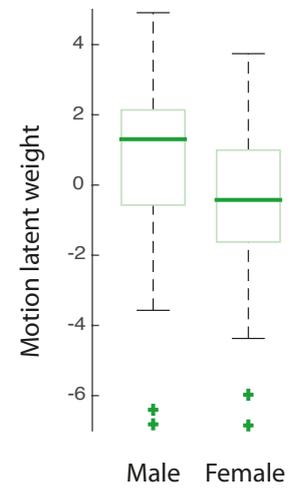
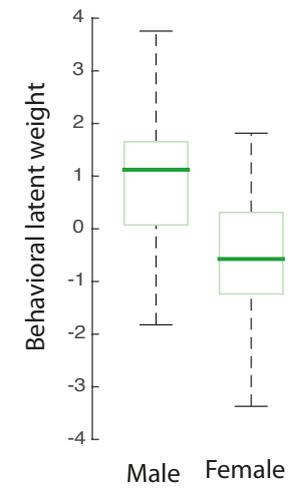

**Supplementary Table 1**: List of considered behavioral scores. For each score, we report (from left to right in the table) whether it was included in the analysis or not (first column), the reason if not included (1: not available upon download, 2: considered irrelevant, 3: discarded at Principal Components construction stage, or 4: discarded at Principal Components quality control stage), the score name, the number of analyzed Principal Components containing the score (in parenthesis, total Principal Components number computed from the score, with 0 denoting single scores directly fed to the analysis), and the percentage of missing entries. Color in the background denotes scores that were jointly included in a Probabilistic PCA.

| Included? | Why? | Name | $P_{NAN}$ | $n_{PC}$ ($n_{TOT}$) | Domain name |
|---|---|---|---|---|---|
| ✗ | 2 | Age_in_Yrs | | | |
| ✗ | 2 | ZygositySR | | | |
| ✗ | 2 | Mother_ID | | | |
| ✗ | 2 | Father_ID | | | |
| ✗ | 4 | Race | 0 | 1 (1) | |
| ✗ | 4 | Ethnicity | 0 | | |
| ✓ | | Handedness | 0 | 0 | |
| ✗ | 4 | SSAGA_Employ | 0 | 1 (1) | |
| ✗ | 4 | SSAGA_Income | 0 | | |
| ✗ | 4 | SSAGA_Educ | 0 | 1 (1) | |
| ✗ | 4 | SSAGA_InSchool | 0 | | |
| ✗ | 2 | SSAGA_MOBorn | | | |
| ✗ | 2 | SSAGA_Rlshp | | | |
| ✓ | | Height | 0.45 | 2 (2) | Body mass index |
| ✓ | | Weight | 0.45 | | |
| ✓ | | BMI | 0.45 | | |
| ✗ | 3 | SSAGA_BMICat | | | |
| ✗ | 3 | SSAGA_BMICatHeaviest | | | |
| ✗ | 2 | Blood_Drawn | | | |
| ✓ | | Hematocrit_1 | 6.25 | 1 (1) | Hematocrit |
| ✓ | | Hematocrit_2 | 9.82 | | |
| ✓ | | BPSystolic | 0.45 | 1 (1) | Blood pressure |
| ✓ | | BPDiastolic | 0.45 | | |
| ✓ | | ThyroidHormone | 21.43 | 0 | Thyroid hormone |
| ✓ | | HbA1C | 20.98 | 0 | HbA1C |
| ✗ | 3 | Hypothyroidism | | | |
| ✗ | 3 | Hypothyroidism_Onset | | | |
| ✗ | 3 | Hyperthyroidism | | | |
| ✗ | 3 | Hyperthyroidism_Onset | | | |
| ✗ | 3 | OtherEndocrn_Prob | | | |
| ✗ | 3 | OtherEndocrine_ProbOnset | | | |
| ✗ | 2 | Menstrual_RegCycles | | | |
| ✗ | 2 | Menstrual_Explain | | | |
| ✗ | 2 | Menstrual_AgeBegan | | | |
| ✗ | 2 | Menstrual_CycleLength | | | |
| ✗ | 2 | Menstrual_DaysSinceLast | | | |
| ✗ | 2 | Menstrual_AgeIrreg | | | |
| ✗ | 2 | Menstrual_AgeStop | | | |
| ✗ | 2 | Menstrual_MonthsSinceStop | | | |
| ✗ | 2 | Menstrual_UsingBirthControl | | | |
| ✗ | 2 | Menstrual_BirthControlCode | | | |
| ✗ | 2 | FamHist_Moth_Scz | | | |
| ✗ | 2 | FamHist_Moth_Dep | | | |
| ✗ | 2 | FamHist_Moth_BP | | | |
| ✗ | 2 | FamHist_Moth_Anx | | | |
| ✗ | 2 | FamHist_Moth_DrgAlc | | | |
| ✗ | 2 | FamHist_Moth_Alz | | | |
| ✗ | 2 | FamHist_Moth_PD | | | |
| ✗ | 2 | FamHist_Moth_TS | | | |
| ✗ | 2 | FamHist_Moth_None | | | |
| ✗ | 2 | FamHist_Fath_Scz | | | |
| ✗ | 2 | FamHist_ Fath _Dep | | | |
| ✗ | 2 | FamHist_ Fath _BP | | | |
| ✗ | 2 | FamHist_ Fath _Anx | | | |
| ✗ | 2 | FamHist_ Fath _DrgAlc | | | |
| ✗ | 2 | FamHist_ Fath _Alz | | | |
| ✗ | 2 | FamHist_ Fath _PD | | | |
| ✗ | 2 | FamHist_ Fath _TS | | | |
| ✗ | 2 | FamHist_ Fath _None | | | |
| ✗ | 4 | MMSE_Score | | | |
| ✗ | 3 | PSQI_Score | | | |
| ✓ | | PSQI_Comp1 | 0 | 1 (6) | Sleep |
| ✓ | | PSQI_Comp2 | 0 | | |
| ✓ | | PSQI_Comp3 | 0 | | |
| ✓ | | PSQI_Comp4 | 0 | | |
| ✓ | | PSQI_Comp5 | 0 | | |
| ✓ | | PSQI_Comp6 | 0 | | |
| ✓ | | PSQI_Comp7 | 0 | | |
| ✓ | | PicSeq_Unadj | 0 | 1 (1) | Episodic memory |
| ✓ | | PicSeq_AgeAdj | 0 | | |
| ✓ | | CardSort_Unadj | 0.45 | 1 (1) | Cognitive flexibility |
| ✓ | | CardSort_AgeAdj | 0.45 | | |
| ✓ | | Flanker_Unadj | 0 | 1 (1) | Inhibition |
| ✓ | | Flanker_AgeAdj | 0 | | |
| ✓ | | PMAT24_A_CR | 1.79 | 1 (2) | Fluid intelligence |
| ✓ | | PMAT24_A_SI | 1.79 | | |
| ✓ | | PMAT24_A_RTCR | 1.79 | | |
| ✓ | | ReadEng_Unadj | 0 | 1 (1) | Reading decoding |
| ✓ | | ReadEng_AgeAdj | 0 | | |
| ✓ | | PicVocab_Unadj | 0 | 1 (1) | Vocabulary comprehension |
| ✓ | | PicVocab_AgeAdj | 0 | | |
| ✓ | | ProcSpeed_Unadj | 0 | 1 (1) | Processing speed |
| ✓ | | ProcSpeed_AgeAdj | 0 | | |
| ✗ | 3 | DDisc_SV_1mo_200 | | | |
| ✗ | 3 | DDisc_SV_6mo_200 | | | |
| ✗ | 3 | DDisc_SV_1yr_200 | | | |
| ✗ | 3 | DDisc_SV_3yr_200 | | | |
| ✗ | 3 | DDisc_SV_5yr_200 | | | |
| ✗ | 3 | DDisc_SV_10r_200 | | | |
| ✗ | 3 | DDisc_SV_1mo_40k | | | |
| ✗ | 3 | DDisc_SV_6mo_40k | | | |
| ✗ | 3 | DDisc_SV_1yr_40k | | | |
| ✗ | 3 | DDisc_SV_3yr_40k | | | |
| ✗ | 3 | DDisc_SV_5yr_40k | | | |
| ✗ | 3 | DDisc_SV_10yr_40k | | | |
| ✓ | | DDisc_AUC_200 | 1.79 | 1 (1) | Impulsivity |
| ✓ | | DDisc_AUC_40k | 1.79 | | |
| ✓ | | VSPLOT_TC | 1.79 | 2 (2) | Spatial orientation |
| ✓ | | VSPLOT_CRTE | 1.79 | | |
| ✓ | | VSPLOT_TC | 1.79 | | |
| ✓ | | SCPT_TP | 1.79 | 3 (4) | Sustained attention |
| ✓ | | SCPT_TN | 1.79 | | |
| ✓ | | SCPT_FP | 1.79 | | |
| ✓ | | SCPT_FN | 1.79 | | |
| ✓ | | SCPT_TPRT | 2.32 | | |
| ✓ | | SCPT_SEN | 1.79 | | |
| ✓ | | SCPT_SPEC | 1.79 | | |
| ✓ | | SCPT_LRNR | 1.79 | | |
| ✓ | | IWRD_TOT | 1.79 | 1 (1) | Verbal episodic memory |
| ✓ | | IWRD_RTC | 1.79 | | |
| ✓ | | ListSort_Unadj | 0 | 1 (1) | Working memory |
| ✓ | | ListSort_AgeAdj | 0 | | |
| ✓ | | ER40_CR | 1.79 | 1 (1) | Emotion recognition (global) |
| ✓ | | ER40_CRT | 1.79 | | |
| ✓ | | ER40ANG | 1.79 | 1 (4) | Emotion recognition (specific) |
| ✓ | | ER40FEAR | 1.79 | | |
| ✓ | | ER40HAP | 1.79 | | |
| ✓ | | ER40NOE | 1.79 | | |
| ✓ | | ER40SAD | 1.79 | | |
| ✓ | | AngAffect_Unadj | 0 | 1 (5) | Negative affect |
| ✓ | | AngHostil_Unadj | 0 | | |
| ✓ | | AngAggr_Unadj | 0 | | |
| ✓ | | FearAffect_Unadj | 0 | | |
| ✓ | | FearSomat_Unadj | 0 | | |
| ✓ | | Sadness_Unadj | 0 | | |
| ✓ | | LifeSatisf_Unadj | 0 | 1 (2) | Psychological well-being |
| ✓ | | MeanPurp_Unadj | 0 | | |
| ✓ | | PosAffect_Unadj | 0 | | |
| ✓ | | Friendship_Unadj | 0 | 1 (5) | Social relationships |
| ✓ | | Loneliness_Unadj | 0 | | |
| ✓ | | PercHostil_Unadj | 0 | | |
| ✓ | | PercReject_Unadj | 0 | | |
| ✓ | | EmotSupp_Unadj | 0 | | |
| ✓ | | IntruSupp_Unadj | 0 | | |
| ✓ | | PercStress_Unadj | 0 | 1 (1) | Stress/Self-efficacy |
| ✓ | | SelfEff_Unadj | 0 | | |
| ✓ | | Endurance_Unadj | 0.45 | 1 (1) | Endurance |
| ✓ | | Endurance_AgeAdj | 0.45 | | |
| ✓ | | GaitSpeed_Comp | 0 | 0 | Locomotion |
| ✓ | | Dexterity_Unadj | 0 | 1 (1) | Dexterity |
| ✓ | | Dexterity_AgeAdj | 0 | | |
| ✓ | | Strength_Unadj | 0 | 1 (1) | Strength |
| ✓ | | Strength_AgeAdj | 0 | | |
| ✓ | | NEOFAC_A | 1.34 | 2 (4) | Personality |
| ✓ | | NEOFAC_O | 1.34 | | |
| ✓ | | NEOFAC_C | 1.34 | | |
| ✓ | | NEOFAC_N | 1.34 | | |
| ✓ | | NEOFAC_E | 1.34 | | |
| ✓ | | ASR_Anxd_Raw | 1.34 | 1 (3) | Anxiety |
| ✓ | | ASR_Anxd_Pct | 1.34 | | |
| ✓ | | DSM_Anxi_Raw | 1.34 | | |
| ✓ | | DSM_Anxi_T | 1.34 | | |
| ✓ | | ASR_Witd_Raw | 1.34 | 1 (1) | Withdrawal |
| ✓ | | ASR_Witd_Pct | 1.34 | | |
| ✓ | | ASR_Soma_Raw | 1.34 | 1 (3) | Somatic problems |
| ✓ | | ASR_Soma_Pct | 1.34 | | |
| ✓ | | DSM_Somp_Raw | 1.34 | | |
| ✓ | | DSM_Somp_T | 1.34 | | |
| ✓ | | ASR_Thot_Raw | 1.34 | 1 (1) | Thought problems |
| ✓ | | ASR_Thot_Pct | 1.34 | | |
| ✓ | | ASR_Attn_Raw | 1.34 | 1 (1) | Attention problems |
| ✓ | | ASR_Attn_Pct | 1.34 | | |
| ✓ | | ASR_Aggr_Raw | 1.34 | 1 (1) | Aggressive behavior |
| ✓ | | ASR_Aggr_Pct | 1.34 | | |
| ✓ | | ASR_Rule_Raw | 1.34 | 1 (1) | Rule breaking |
| ✓ | | ASR_Rule_Pct | 1.34 | | |
| ✓ | | ASR_Intr_Raw | 1.34 | 1 (1) | Intrusiveness |
| ✓ | | ASR_Intr_Pct | 1.34 | | |
| ✗ | 2 | ASR_Oth_Raw | | | |
| ✗ | 2 | ASR_Crit_Raw | | | |
| ✓ | | ASR_Intn_Raw | 1.34 | 1 (1) | Internalization |
| ✓ | | ASR_Intn_T | 1.34 | | |
| ✓ | | ASR_Extn_Raw | 1.34 | 1 (1) | Externalization |
| ✓ | | ASR_Extn_T | 1.34 | | |
| ✗ | 2 | ASR_TAO_Sum | | | |
| ✗ | 2 | ASR_Totp_Raw | | | |
| ✗ | 2 | ASR_Totp_T | | | |
| ✗ | 4 | DSM_Depr_Raw | 1.34 | 1 (3) | |
| ✗ | 4 | DSM_Depr_T | 1.34 | | |
| ✗ | 4 | SSAGA_Depressive_Ep | 1.34 | | |
| ✗ | 4 | SSAGA_Depressive_Sx | 1.34 | | |
| ✓ | | DSM_Avoid_Raw | 1.34 | 1 (1) | Avoidance |
| ✓ | | DSM_Avoid_T | 1.34 | | |
| ✓ | | DSM_Adh_Raw | 1.34 | 1 (1) | |
| ✓ | | DSM_Adh_T | 1.34 | | |
| ✓ | | DSM_Inat_Raw | | 0 | Inattention |
| ✓ | | DSM_Hype_Raw | | 0 | Hyperactivity |
| ✓ | | DSM_Antis_Raw | 1.34 | 1 (1) | Antisocial behavior |
| ✓ | | DSM_Antis_T | 1.34 | | |
| ✗ | 2 | SSAGA_ChildhoodConduct | | | |
| ✗ | 2 | SSAGA_PanicDisorder | | | |
| ✗ | 2 | SSAGA_Agoraphobia | | | |
| ✓ | | Noise_Comp | 0.89 | 0 | Audition |
| ✓ | | Odor_Unadj | 0.45 | 1 (1) | Olfaction |
| ✓ | | Odor_AgeAdj | 0.45 | | |
| ✓ | | PainIntens_RawScore | 0 | 1 (1) | Pain intensity |
| ✓ | | PainInterf_Tscore | 0 | | |
| ✓ | | Taste_Unadj | 0.45 | 1 (1) | Taste |
| ✓ | | Taste_AgeAdj | 0.45 | | |
| ✗ | 2 | Color_Vision | | | |
| ✗ | 2 | Eye | | | |
| ✗ | 2 | EVA_Num | | | |
| ✗ | 2 | EVA_Denom | | | |
| ✗ | 2 | Correction | | | |
| ✓ | | Mars_Log_Score | 1.79 | 2 (2) | Contrast sensitivity |
| ✓ | | Mars_Errs | 2.23 | | |
| ✓ | | Mars_Final | 2.23 | | |
| ✗ | 1 | Breathalyzer_Over_05 | | | |
| ✗ | 1 | Breathalyzer_Over_08 | | | |
| ✗ | 1 | Cocaine | | | |
| ✗ | 1 | THC | | | |
| ✗ | 1 | Opiates | | | |
| ✗ | 1 | Amphetamines | | | |
| ✗ | 1 | MethAmphetamine | | | |
| ✗ | 1 | Oxycontin | | | |
| ✓ | | Total_Drinks_7days | 2.23 | 1 (6) | Alcohol consumption |
| ✓ | | Num_Days_Drank_7days | 0 | | |
| ✓ | | SSAGA_Alc_D4_Dp_Sx | 0 | | |
| ✓ | | SSAGA_Alc_D4_Ab_Sx | 0 | | |
| ✓ | | SSAGA_Alc_12_Drinks_Per_Day | 4.91 | | |
| ✓ | | SSAGA_Alc_12_Frq | 4.91 | | |
| ✓ | | SSAGA_Alc_12_Frq_Drk | 4.91 | | |
| ✗ | 3 | Avg_Weekday_Drinks_7days | | | |
| ✗ | 3 | Avg_Weekend_Drinks_7days | | | |
| ✗ | 3 | Total_Beer_Wine_Cooler_7days | | | |
| ✗ | 3 | Avg_Weekday_Beer_Wine_Cooler_7days | | | |
| ✗ | 3 | Avg_Weekend_Beer_Wine_Cooler_7days | | | |
| ✗ | 3 | Total_Malt_Liquor_7days | | | |
| ✗ | 3 | Avg_Weekday_Malt_Liquor_7days | | | |
| ✗ | 3 | Avg_Weekend_Malt_Liquor_7days | | | |
| ✗ | 3 | Total_Wine_7days | | | |
| ✗ | 3 | Avg_Weekday_Wine_7days | | | |
| ✗ | 3 | Avg_Weekend_Wine_7days | | | |
| ✗ | 3 | Total_Hard_Liquor_7days | | | |
| ✗ | 3 | Avg_Weekday_Hard_Liquor_7days | | | |
| ✗ | 3 | Avg_Weekend_Hard_Liquor_7days | | | |
| ✗ | 3 | Total_Other_Alc_7days | | | |
| ✗ | 3 | Avg_Weekday_Other_Alc7days | | | |
| ✗ | 3 | Avg_Weekend_Other_Alc_7days | | | |
| ✗ | 3 | SSAGA_Alc_D4_Ab_Dx | | | |
| ✗ | 3 | SSAGA_Alc_D4_Dp_Dx | | | |
| ✗ | 3 | SSAGA_Acl_12_Frq_5plus | | | |
| ✗ | 3 | SSAGA_Alc_12_Max_Drinks | | | |
| ✗ | 3 | SSAGA_Alc_Age_1st_Use | | | |
| ✗ | 3 | SSAGA_Alc_Hvy_Drinks_Per_Day | | | |
| ✗ | 3 | SSAGA_Alc_Hvy_Frq | | | |
| ✗ | 3 | SSAGA_Alc_Hvy_Frq_5plus | | | |
| ✗ | 3 | SSAGA_Alc_Hvy_Frq_Drk | | | |
| ✗ | 3 | SSAGA_Alc_Hvy_Max_Drinks | | | |
| ✓ | | Total_Any_Tobacco_7days | 2.23 | 1 (3) | Tobacco consumption |
| ✓ | | Times_Used_Any_Tobacco_Today | 2.23 | | |
| ✓ | | Num_Days_Used_Any_Tobacco_7days | 0 | | |
| ✓ | | SSAGA_TB_Smoking_History | 0 | | |
| ✗ | 3 | Avg_Weekday_Any_Tobacco_7days | | | |
| ✗ | 3 | Avg_Weekend_Any_Tobacco_7days | | | |
| ✗ | 3 | Total_Cigarettes_7days | | | |
| ✗ | 3 | Avg_Weekday_Cigarettes_7days | | | |
| ✗ | 3 | Avg_Weekend_Cigarettes_7days | | | |
| ✗ | 3 | Total_Cigars_7days | | | |
| ✗ | 3 | Avg_Weekday_Cigars_7days | | | |
| ✗ | 3 | Avg_Weekend_Cigars_7days | | | |
| ✗ | 3 | Total_Pipes_7days | | | |
| ✗ | 3 | Avg_Weekday_Pipes_7days | | | |
| ✗ | 3 | Avg_Weekend_Pipes_7days | | | |
| ✗ | 3 | Total_Chew_7days | | | |
| ✗ | 3 | Avg_Weekday_Chew_7days | | | |
| ✗ | 3 | Avg_Weekend_Chew_7days | | | |
| ✗ | 3 | Total_Snuff_7days | | | |
| ✗ | 3 | Avg_Weekday_Snuff_7days | | | |
| ✗ | 3 | Avg_Weekend_Snuff_7days | | | |
| ✗ | 3 | Total_Other_Tobacco_7days | | | |
| ✗ | 3 | Avg_Weekday_Other_Tobacco_7days | | | |
| ✗ | 3 | Avg_Weekend_Other_Tobacco_7days | | | |
| ✗ | 2 | SSAGA_FTND_Score | | | |
| ✗ | 2 | SSAGA_HSI_Score | | | |
| ✗ | 2 | SSAGA_TB_Age_1st_Cig | | | |
| ✗ | 2 | SSAGA_TB_DSM_Difficulty_Quitting | | | |
| ✗ | 2 | SSAGA_TB_DSM_Tolerance | | | |
| ✗ | 2 | SSAGA_TB_DSM_Withdrawal | | | |
| ✗ | 2 | SSAGA_TB_Hvy_CPD | | | |
| ✗ | 2 | SSAGA_TB_Max_Cigs | | | |
| ✗ | 2 | SSAGA_TB_Reg_CPD | | | |
| ✗ | 2 | SSAGA_TB_Still_Smoking | | | |
| ✗ | 2 | SSAGA_TB_Yrs_Since_Quit | | | |
| ✗ | 2 | SSAGA_TB_Yrs_Smoked | | | |
| ✓ | | SSAGA_Times_Used_Illicits | 0 | 1 (6) | Drugs consumption |
| ✓ | | SSAGA_Times_Used_Cocaine | 0 | | |
| ✓ | | SSAGA_Times_Used_Hallucinogens | 0 | | |
| ✓ | | SSAGA_Times_Used_Opiates | 0 | | |
| ✓ | | SSAGA_Times_Used_Sedatives | 0 | | |
| ✓ | | SSAGA_Times_Used_Stimulants | 0 | | |
| ✓ | | SSAGA_Mj_Times_Used | 0 | | |
| ✗ | 2 | SSAGA_Mj_Use | | | |
| ✗ | 2 | SSAGA_Mj_Ab_Dep | | | |
| ✗ | 2 | SSAGA_Mj_Age_1st_Use | | | |
| ✓ | | Language_Task_Story_Acc | 5.36 | 2 (5) | Language task |
| ✓ | | Language_Task_Story_Median_RT | 5.36 | | |
| ✓ | | Language_Task_Story_Avg_Difficulty_Level | 5.36 | | |
| ✓ | | Language_Task_Math_Acc | 5.36 | | |
| ✓ | | Language_Task_Math_Median_RT | 5.36 | | |
| ✓ | | Language_Task_Math_Avg_Difficulty_Level | 5.36 | | |
| ✓ | | Emotion_Task_Acc | 5.36 | 2 (5) | Emotion task |
| ✓ | | Emotion_Task_Median_RT | 5.36 | | |
| ✓ | | Emotion_Task_Face_Acc | 5.36 | | |
| ✓ | | Emotion_Task_Face_Median_RT | 5.36 | | |
| ✓ | | Emotion_Task_Shape_Acc | 5.36 | | |
| ✓ | | Emotion_Task_Shape_Median_RT | 5.36 | | |
| ✗ | 3 | Gambling_Task_Perc_Larger | | | |
| ✗ | 3 | Gambling_Task_Perc_Smaller | | | |
| ✗ | 3 | Gambling_Task_Perc_NLR | | | |
| ✗ | 3 | Gambling_Task_Median_RT_Larger | | | |
| ✗ | 3 | Gambling_Task_Median_RT_Smaller | | | |
| ✓ | | Gambling_Task_Reward_Perc_Larger | 0.45 | 3 (6) | Gambling task |
| ✓ | | Gambling_Task_Reward_Median_RT_Larger | 1.34 | | |
| ✓ | | Gambling_Task_Reward_Perc_Smaller | 0.45 | | |
| ✓ | | Gambling_Task_Reward_Median_RT_Smaller | 2.23 | | |
| ✓ | | Gambling_Task_Punish_Perc_Larger | 0.45 | | |
| ✓ | | Gambling_Task_Punish_Median_RT_Larger | 0.45 | | |
| ✓ | | Gambling_Task_Punish_Perc_Smaller | 0.45 | | |
| ✓ | | Gambling_Task_Punish_Median_RT_Smaller | 0.89 | | |
| ✗ | 3 | Gambling_Task_Reward_Perc_NLR | | | |
| ✗ | 3 | Gambling_Task_Punish_Perc_NLR | | | |
| ✗ | 3 | Relational_Task_Acc | | | |
| ✗ | 3 | Relational_Task_Median_RT | | | |
| ✓ | | Relational_Task_Match_Acc | 6.25 | 2 (3) | Relational task |
| ✓ | | Relational_Task_Match_Median_RT | 6.25 | | |
| ✓ | | Relational_Task_Rel_Acc | 6.25 | | |
| ✓ | | Relational_Task_Rel_Median_RT | 6.25 | | |
| ✗ | 3 | Social_Task_Perc_Random | | | |
| ✗ | 3 | Social_Task_Perc_TOM | | | |
| ✗ | 3 | Social_Task_Perc_Unsure | | | |
| ✗ | 3 | Social_Task_Perc_NLR | | | |
| ✗ | 3 | Social_Task_Median_RT_Random | | | |
| ✗ | 3 | Social_Task_Median_RT_TOM | | | |
| ✗ | 3 | Social_Task_Median_RT_Unsure | | | |
| ✓ | | Social_Task_Random_Perc_Random | 5.36 | 1 (5) | Social task |
| ✓ | | Social_Task_Random_Perc_TOM | 5.36 | | |
| ✓ | | Social_Task_Random_Median_RT_Random | 5.36 | | |
| ✓ | | Social_Task_TOM_Perc_Random | 5.36 | | |
| ✓ | | Social_Task_TOM_Perc_TOM | 5.36 | | |
| ✓ | | Social_Task_TOM_Median_RT_TOM | 5.36 | | |
| ✗ | 3 | Social_Task_Random_Perc_Unsure | | | |
| ✗ | 3 | Social_Task_Random_Perc_NLR | | | |
| ✗ | 3 | Social_Task_Random_Median_RT_TOM | | | |
| ✗ | 3 | Social_Task_Random_M | | | |

| | | edian_RT_Unsure | | | |
|---|---|---|---|---|---|
| ✗ | 3 | Social_Task_TOM_Perc_Unsure | | | |
| ✗ | 3 | Social_Task_TOM_Perc_NLR | | | |
| ✗ | 3 | Social_Task_TOM_Median_RT_Random | | | |
| ✗ | 3 | Social_Task_TOM_Median_RT_Unsure | | | |
| ✗ | 3 | WM_Task_Acc | | | |
| ✗ | 3 | WM_Task_Median_RT | | | |
| ✓ | | WM_Task_2bk_Acc | 1.34 | 1 (3) | Working memory task |
| ✓ | | WM_Task_2bk_Median_RT | 1.34 | | |
| ✓ | | WM_Task_0bk_Acc | 0.45 | | |
| ✓ | | WM_Task_0bk_Median_RT | 1.34 | | |
| ✗ | 3 | WM_Task_0bk_Body_Acc | | | |
| ✗ | 3 | WM_Task_0bk_Body_Acc_Target | | | |
| ✗ | 3 | WM_Task_0bk_Body_Acc_Nontarget | | | |
| ✗ | 3 | WM_Task_0bk_Face_Acc | | | |
| ✗ | 3 | WM_Task_0bk_Face_Acc_Target | | | |
| ✗ | 3 | WM_Task_0bk_Face_ACC_Nontarget | | | |
| ✗ | 3 | WM_Task_0bk_Place_Acc | | | |
| ✗ | 3 | WM_Task_0bk_Place_Acc_Target | | | |
| ✗ | 3 | WM_Task_0bk_Place_Acc_Nontarget | | | |
| ✗ | 3 | WM_Task_0bk_Tool_Acc | | | |
| ✗ | 3 | WM_Task_0bk_Tool_Acc_Target | | | |
| ✗ | 3 | WM_Task_0bk_Tool_Acc_Nontarget | | | |
| ✗ | 3 | WM_Task_2bk_Body_Acc | | | |
| ✗ | 3 | WM_Task_2bk_Body_Acc_Target | | | |
| ✗ | 3 | WM_Task_2bk_Body_Acc_Nontarget | | | |
| ✗ | 3 | WM_Task_2bk_Face_Acc | | | |
| ✗ | 3 | WM_Task_2bk_Face_Acc_Target | | | |
| ✗ | 3 | WM_Task_2bk_Face_ACC_Nontarget | | | |
| ✗ | 3 | WM_Task_2bk_Place_Acc | | | |
| ✗ | 3 | WM_Task_2bk_Place_Acc_Target | | | |
| ✗ | 3 | WM_Task_2bk_Place_Ac | | | |
| ✗ | | c_Nontarget | | | |
| ✗ | 3 | WM_Task_2bk_Tool_Acc | | | |
| ✗ | 3 | WM_Task_2bk_Tool_Acc_Target | | | |
| ✗ | 3 | WM_Task_2bk_Tool_Acc_Nontarget | | | |
| ✗ | 3 | WM_Task_0bk_Body_Median_RT | | | |
| ✗ | 3 | WM_Task_0bk_Body_Median_RT_Target | | | |
| ✗ | 3 | WM_Task_0bk_Body_Median_RT_Nontarget | | | |
| ✗ | 3 | WM_Task_0bk_Face_Median_RT | | | |
| ✗ | 3 | WM_Task_0bk_Face_Median_RT_Target | | | |
| ✗ | 3 | WM_Task_0bk_Face_Median_RT_Nontarget | | | |
| ✗ | 3 | WM_Task_0bk_Place_Median_RT | | | |
| ✗ | 3 | WM_Task_0bk_Place_Median_RT_Target | | | |
| ✗ | 3 | WM_Task_0bk_Place_Median_RT_Nontarget | | | |
| ✗ | 3 | WM_Task_0bk_Tool_Median_RT | | | |
| ✗ | 3 | WM_Task_0bk_Tool_Median_RT_Target | | | |
| ✗ | 3 | WM_Task_0bk_Tool_Median_RT_Nontarget | | | |
| ✗ | 3 | WM_Task_2bk_Body_Median_RT | | | |
| ✗ | 3 | WM_Task_2bk_Body_Median_RT_Target | | | |
| ✗ | 3 | WM_Task_2bk_Body_Median_RT_Nontarget | | | |
| ✗ | 3 | WM_Task_2bk_Face_Median_RT | | | |
| ✗ | 3 | WM_Task_2bk_Face_Median_RT_Target | | | |
| ✗ | 3 | WM_Task_2bk_Face_Median_RT_Nontarget | | | |
| ✗ | 3 | WM_Task_2bk_Place_Median_RT | | | |
| ✗ | 3 | WM_Task_2bk_Place_Median_RT_Target | | | |
| ✗ | 3 | WM_Task_2bk_Place_Median_RT_Nontarget | | | |
| ✗ | 3 | WM_Task_2bk_Tool_Median_RT | | | |
| ✗ | 3 | WM_Task_2bk_Tool_Median_RT_Target | | | |
| ✗ | 3 | WM_Task_2bk_Tool_Median_RT_Nontarget | | | |

| X | $G_1$ | $G_2$ | $G_3$ | $G_4$ |
|---|---|---|---|---|
| $G_1$ |  | 1.14 | -8.64 | -3.09 |
| $G_2$ | 0.26 |  | -8.25 | -3.66 |
| $G_3$ | 0 | 0 |  | 5.93 |
| $G_4$ | 0.002 | 0.0003 | 0 |  |

| Y | $G_1$ | $G_2$ | $G_3$ | $G_4$ |
|---|---|---|---|---|
| $G_1$ |  | -0.25 | -6.36 | -1.94 |
| $G_2$ | 0.8 |  | -5.74 | -1.55 |
| $G_3$ | 0 | 0 |  | 4.31 |
| $G_4$ | 0.05 | 0.12 | 0 |  |

| Z | $G_1$ | $G_2$ | $G_3$ | $G_4$ |
|---|---|---|---|---|
| $G_1$ |  | -3.27 | -8.38 | -2.85 |
| $G_2$ | 0.0011 |  | -6.13 | 0.7 |
| $G_3$ | 0 | 0 |  | 6.07 |
| $G_4$ | 0.004 | 0.48 | 0 |  |

| α | $G_1$ | $G_2$ | $G_3$ | $G_4$ |
|---|---|---|---|---|
| $G_1$ |  | -8.31 | -8.85 | -4.48 |
| $G_2$ | 0 |  | -2.04 | 4.28 |
| $G_3$ | 0 | 0.04 |  | 5.07 |
| $G_4$ | 0 | 0 | 0 |  |

| β | $G_1$ | $G_2$ | $G_3$ | $G_4$ |
|---|---|---|---|---|
| $G_1$ |  | -6.21 | -8.8 | -4.63 |
| $G_2$ | 0 |  | -4.91 | 1.8 |
| $G_3$ | 0 | 0 |  | 5.83 |
| $G_4$ | 0 | 0.07 | 0 |  |

| γ | $G_1$ | $G_2$ | $G_3$ | $G_4$ |
|---|---|---|---|---|
| $G_1$ |  | 1.93 | -7.69 | -8.24 |
| $G_2$ | 0.05 |  | -7.94 | -7.91 |
| $G_3$ | 0 | 0 |  | -4.05 |
| $G_4$ | 0 | 0 | 0.0001 |  |

**Supplementary Table 2**: For all 6 motion parameters (each of the 6 sub-tables), statistics for pair-wise comparison of average motion over time. The upper right diagonals contain differences in rank-sum statistics (a negative entry means that the group denoted in the column showed a larger statistic), while the lower left diagonals display the related p-values.

## Supplementary Material

This Supplementary Material is subdivided into three sections: in the first one, we give extended explanations on spectral clustering and Partial Least Square analysis, two approaches used in our work. In the second section, we provide details on the exact constituents of each of the 45 behavioral domain scores that we analyzed. In the third section, we assess the robustness of classification results and motion/behavioral saliences to changes in scrubbing settings.

### Section 1: Analytical Details

**Graph theory, spectral clustering and consensus clustering**

There exist many ways to cluster a dataset into a subset of distinct groups. In our case, we applied *spectral clustering* on our data matrix $X \in \mathbb{R}^{224 \times 36}$. Each of our data points can thus be expressed as a vector $\boldsymbol{x_i} \in \mathbb{R}^{36 \times 1}$, arranged as the rows of $X$.

Spectral clustering necessitates (1) the definition of a graph summarizing the data at hand, (2) the extraction of meaningful components summarizing the data based on the graph architecture (which can be understood as a dimensionality reduction approach), and (3) classification using the extracted components. We go through those three steps in details below.

*Graph definition*

Let us consider a graph $G \triangleq (\mathcal{V}, \mathcal{E})$, where $\mathcal{V}$ is its constituting set of nodes and $\mathcal{E}$ the set of edges linking those nodes. We denote by $w_{i,j}$ the edge weight between nodes $i$ and $j$; a larger value indicates a closer similarity between the nodes.

Graphs can be used to represent a wide array of systems, such as transportation networks, metabolic networks or social networks. In neuroscience, a classical approach has been to define nodes as brain regions, and edges as their structural or functional connectivity.

Here, we adopt another approach in which we define the nodes as the 224 subjects considered in our analyses. Edge weights were set using an N-nearest neighbor criterion, in which each node was linked to its N closest neighbors (as quantified from cosine similarity) only. Let $\mathcal{N}_i$ the N-neighborhood of node $i$; edge weights were initialized as:

$$w_{i,j} = \begin{cases} e^{-\frac{d_{i,j}^2}{\sigma_i^2}} & \text{if } j \in \mathcal{N}_i, \\ 0 & \text{else} \end{cases}$$

where $d_{i,j} = 1 - \frac{x_i \cdot x_j}{||x_i||\,||x_j||}$ is the cosine distance between data points $i$ and $j$, and $\sigma_i$ is the average of all distance between $i$ and its N nearest neighbors. Weights can take values between 0 (infinitely distant/non-neighboring data points) and 1 (identical data points).

Edge weights can be efficiently summarized into the *adjacency matrix A* of the system.

***Dimensionality reduction***

We first define the symmetric, positive-definite Laplacian matrix of the system as $L = D - A$, where $D$ is a diagonal matrix containing nodal degrees. The nodal degree of node $i$ is the sum of incoming edges: $d_i = \sum_{j \in \mathcal{N}_i} w_{i,j}$. In our analyses, we considered the normalized version of the Laplacian, given by:

$$L_N = D^{-0.5} L\, D^{-0.5}.$$

$L_N$ can equivalently be expressed as an eigenvalue decomposition, as $L_N = U \Sigma U^T$. In this expression, $U$ is the matrix of eigenvectors (arranged in columns) and $\Sigma$ is a diagonal matrix containing the associated eigenvalues in its diagonal. We consider sorted eigenvalue/eigenvector pairs in decreasing eigenvalue order.

The first three eigenvectors with non-null eigenvalue happen to be an optimal basis for classification; expressing our 36-dimensional data points in this three-dimensional space thus operates as a nonlinear dimensionality reduction approach.

***Clustering***

To partition the data into clusters, k-means clustering is performed on the 224 x 3 dimensionally reduced dataset. To select the optimal number of clusters, we used consensus clustering (Monti et al. 2003), a subsampling-based assessment of robustness.

In more details, the clustering process was repeatedly run (100 times) over 80% of the data points (*i.e.*, 179 subjects), for cluster numbers $k$ ranging from 2 to 17. For each $k$, a consensus matrix summarizing how frequently two data points would be clustered together was derived. Since the goal in a good clustering scheme is to either always cluster two data points together, or to never do so, the goal is to find a $k$ for which the proportion of ambiguously clustered pairs (PAC; Şenbabaoğlu et al. 2014), linked to intermediate consensus values, is the lowest. As showed in **Figure 1A**, $k = 4$ stood as a clear optimum.

**Partial Least Square (PLS) Analysis**

PLS is a multivariate approach that enables to extract co-varying components between two types of measures. In what follows, we will be considering the matrix of spatiotemporal motion features $X \in \mathbb{R}^{36 \times 224}$, and the matrix of behavioral domain scores $Y \in \mathbb{R}^{45 \times 224}$.

The goal in PLS analysis is to extract covariance components from the data. To do so, we first consider the covariance matrix between spatiotemporal motion features and behavioral domain scores:

$$R = Y \cdot X \in \mathbb{R}^{45 \times 36}.$$

This matrix can be equivalently expressed in the form of a singular value decomposition:

$$R = U\Sigma V^T.$$

In the above equation, the matrix $U$ contains the left singular vectors of $R$, arranged in successive columns. Those vectors form an orthonormal basis; *i.e.*, $U^T \cdot U = I$. The same property applies to the right singular vectors, arranged in the columns of the matrix $V$. As for $\Sigma$, it is a diagonal matrix containing the singular values $\sigma_i, i = 1,\ldots,36$ as its diagonal elements. We assume here that singular vectors and singular values are sorted in decreasing singular value order.

The intuition behind this decomposition is that the full covariance between both datasets is expressed as a weighted low-rank approximation:

$$R = \sum_{i=1}^{36} \sigma_i \boldsymbol{u}_i \boldsymbol{v}_i^T,$$

where $\boldsymbol{u}_i$ and $\boldsymbol{v}_i$ are the i[th] left and right singular vectors, respectively. Since $U$ and $V$ can both be seen as orthonormal bases, it follows that the strength of expression of the components in the investigated pool of subjects can be simply expressed as a projection:

$$L_Y = U^T Y \text{ and } L_X = V^T X,$$

with $L_Y$ denoting the strength of expression of behavioral domain scores, or *behavioral latent weights*, across subjects, and $L_X$ that of spatiotemporal motion features (called *motion latent weights*). The i[th] column of $L_Y$ or $L_X$ contains the weights associated to the i[th] component, while the j[th] row contains all weights associated to subject j.

## Generation of behavioral domain scores

In this work, we describe the relationships between spatiotemporal motion features and a set of 45 domain scores. Each of those scores is obtained as a weighted combination from original scores provided by the HCP. Below, we describe how we selected the original scores to use, and how we combined them. We also provide an interpretation of each domain scores.

### Selection of measures of interest

We chose not include some types of HCP scores into our analysis, because they highlighted family relationships or attributes falling beyond the scope of the present work, or were available in a too limited fraction of subjects. This included:

1. Family relationships between subjects and twin status
2. Psychiatric history of the mother or father
3. Scores reflective of the menstrual cycle (only available in female subjects).

In several cases, *Age adjusted* and *Unadjusted* scores are provided. Both were kept and aggregated in further processing steps, as described below.

Because some domains included many more scores than others, and in order to conduct a balanced analysis, we converted the scores of each domain into only one value through Probabilistic Principal Component Analysis (PPCA; Bishop 1999), which enabled, at the same time, to fill in the few missing entries in each case. Some scores were not retained because they appeared irrelevant to us (labeled '2' in **Supplementary Table 2**; *e.g.*, *has blood been sampled?*), and others because they were considered too specific (that is, would induce overfitting), or overlapped with others (labeled '3').

Eventually, we only retained the domain scores that were sufficiently accurate, according to criteria proposed by Smith et al. (2015); excluded scores at this stage are labeled '4'. Exclusion criteria were:

1. More than 5 unavailable entries (in the case of domain scores that were derived from only one HCP score and did thus not undergo PPCA)
2. The case $\max(\boldsymbol{z}) > 100\bar{z}$, with $\boldsymbol{z} = (\boldsymbol{b} - \boldsymbol{I}\,\text{med}(\boldsymbol{b}))^2$, the vector $b \in \mathbb{R}^{224 \times 1}$ the values for a given score across subjects, and $\boldsymbol{I}$ a unitary vector of appropriate size
3. More than 95% of subjects showing the same score value.

Finally, the remaining domain scores underwent rank-based inverse Gaussian transformation. The final matrix of behavioral information used for the analyses had size 224 x 45 (subjects x domains).

### Derivation of composite domain measures from individual HCP scores

Our analyses relate spatiotemporal motion features to 45 components reflective of anthropometric, behavioral or clinical features from the scanned subjects.

Each of those 45 components is obtained from a set of original scores from the HCP database. Below, for each of the 45 cases, we provide:
1. Related original scores from the HCP database
2. Associated correlation values with the component (Spearman correlation)
3. Resulting interpretation of the component's meaning

**Component 1**
Original HCP scores: Height (0.6), Weight (1.0) and BMI (0.88).

This component mostly reflects the effect of increased weight on BMI; as such, we interpreted it as the detrimental impact on BMI and named the component **Weight**.

**Component 2**
Original HCP scores: Height (0.81), Weight (0.02) and BMI (-0.36).

Larger component values highlight lowered BMI due to greater height of the subjects. As such, we named the component **BMI (good)**.

**Component 3**
Original HCP scores: Systolic blood pressure (0.90) and Diastolic blood pressure (0.92).

This component reflects increases in blood pressure; we named it **Blood pressure**.

**Component 4**
Original HCP scores: PSQI Component 1 (0.76), PSQI Component 2 (0.48), PSQI Component 3 (0.61), PSQI Component 4 (0.66), PSQI Component 5 (0.51), PSQI Component 6 (0.28) and PSQI Component 7 (0.44).

Larger component scores denote more sleep issues in terms of a specific characteristic. Given the uniformly positive correlation patterns, this component reflects global sleep problems; we thus named it **Sleep problems**.

**Component 5**
Original HCP scores: Dimensional Change Card Sort Unadjusted (0.99) and Dimensional Change Card Sort Age Adjusted (0.99).

Dimensional Change Card Sort scores measure cognitive flexibility: thus, we named this component **Cognitive flexibility**.

**Component 6**
Original HCP scores: Penn Progressive Matrices Number of Correct Responses (-0.94), Penn Progressive Matrices Total Skipped Items (0.93), and Penn Progressive Matrices Median Reaction Time for Correct Responses (-0.92).

Penn Progressive Matrices measure fluid intelligence. Larger component values go with worse performance, more skipped items and lower response time for correct responses. As such, it may either reflect lowered fluid intelligence (due to the worsened performance), or the fact that subjects did not care about the task and simply sped through without focusing (excessive carelessness, hence the lowered response time).

We thus named the component **Fluid intelligence (bad)/carelessness**.

**Component 7**
Original HCP scores: Delay Discounting Area Under the Curve for $200 (0.88) and Delay Discounting Area Under the Curve for $40000 (0.94).

Larger scores reflect the fact that subjects attribute greater subjective value to an amount of money if provided with a temporal delay; as such, larger component values highlight greater self-regulation, or lowered impulsivity. We thus named the component **Self-regulation**.

**Component 8**
Original HCP scores: Variable Short Penn Line Orientation Total Number Correct (-0.96), Variable Short Penn Line Orientation Median Response Time Divided by Expected Number of Clicks for Correct (-0.33), and Variable Short Penn Line Orientation Total Positions Off for All Trials (0.97).

Larger component values indicated worsened spatial orientation abilities (less correct responses and more off trials); we thus named the component **Spatial orientation (bad)**.

**Component 9**
Original HCP scores: Variable Short Penn Line Orientation Total Number Correct (-0.23), Variable Short Penn Line Orientation Median Response Time Divided by Expected Number of Clicks for Correct (0.91), and Variable Short Penn Line Orientation Total Positions Off for All Trials (0.19).

This component reflects the time taken to answer in the test more than the actual performance (for which correlation values are milder); we thus named it **Spatial orientation (RT)**.

**Component 10**
Original HCP scores: Short Penn Continuous Performance Test True Positives (-0.7), Short Penn Continuous Performance Test True Negatives (0.0), Short Penn Continuous Performance Test Open Positives (0.0), Short Penn Continuous Performance Test Open Negatives (0.7), Short Penn Continuous Performance Test Median Response Time for True Positive Responses (0.63), Short Penn Continuous Performance Test Sensitivity (-0.7), Short Penn Continuous Performance Test Specificity (0), Short Penn Continuous Performance Test Longest Run of Non-Responses (0.34).

This component shows larger values when the subjects show worse sensitivity (*i.e.*, often miss a stimulus that they should have responded to), and also to a lower degree when subjects show more non-responses or take longer to respond correctly. Thus, it reflects worse sustained attention abilities, and we named it **Sustained attention (non sensitivity)**.

**Component 11**
Original HCP scores: Short Penn Continuous Performance Test True Positives (0.2), Short Penn Continuous Performance Test True Negatives (0.98), Short Penn Continuous Performance Test Open Positives (-0.98), Short Penn Continuous Performance Test Open Negatives (-0.2), Short Penn Continuous Performance Test Median Response Time for True Positive Responses (0.31), Short Penn Continuous Performance Test Sensitivity (0.2), Short Penn Continuous Performance Test Specificity (0.98), Short Penn Continuous Performance Test Longest Run of Non-Responses (0.29).

This component shows larger values when the subjects correctly rule a stimulus out; we thus named it **Sustained attention (specificity)**.

**Component 12**
Original HCP scores: Short Penn Continuous Performance Test True Positives (0.36), Short Penn Continuous Performance Test True Negatives (-0.14), Short Penn Continuous Performance Test Open Positives (0.14), Short Penn Continuous Performance Test Open Negatives (-0.36), Short Penn Continuous Performance Test Median Response Time for True Positive Responses (0.82), Short Penn Continuous Performance Test Sensitivity (0.36), Short Penn Continuous Performance Test Specificity (-0.14), Short Penn Continuous Performance Test Longest Run of Non-Responses (-0.17).

This component mostly shows larger values when the subjects take longer to correctly recognize a hit; it also shows larger values with better sensitivity. Thus, we named it **Sustained attention (RT/carefulness)**.

**Component 13**
Original HCP scores: Penn Word Memory Test Total Number of Correct Responses (-0.84) and Penn Word Memory Test Median Reaction Time for Correct Responses (0.77).

This component shows larger values when the subjects need longer to respond, and respond worse; we thus named it **Verbal episodic memory (bad)**.

**Component 14**
Original HCP scores: Working Memory Task 2-Back Accuracy (-0.73), Working Memory Task 2-Back Median Reaction Time (0.73), Working Memory Task 0-Back Accuracy (-0.64) and Working Memory Task 0-Back Median Reaction Time (0.82).

This component had larger values for subjects answering worse and needing more time to answer; thus, we named it **Working memory (bad)**.

**Component 15**
Original HCP scores: Gambling Task Reward Percentage Larger (0.09), Gambling Task Reward Median Reaction Time Larger (0.92), Gambling Task Reward Percentage Smaller (-0.09), Gambling Task Reward Median Reaction Time Smaller (0.93), Gambling Task Punish Percentage Larger (-0.07), Gambling Task

Punish Median Reaction Time Larger (0.91), Gambling Task Punish Percentage Smaller (0.07), Gambling Task Punish Median Reaction Time Smaller (0.92).

This component had larger values for subjects who needed more time to respond, regardless of the condition. Thus, we named it **Gambling (RT)**.

**Component 16**
Original HCP scores: Gambling Task Reward Percentage Larger (0.76), Gambling Task Reward Median Reaction Time Larger (0.02), Gambling Task Reward Percentage Smaller (-0.76), Gambling Task Reward Median Reaction Time Smaller (0.02), Gambling Task Punish Percentage Larger (0.85), Gambling Task Punish Median Reaction Time Larger (-0.02), Gambling Task Punish Percentage Smaller (-0.85), Gambling Task Punish Median Reaction Time Smaller (0.02).

This component shows greater values for the subjects who gamble more regardless of whether they will win or lose (the Reward and Punish categories, respectively); thus, we named it **Gambling (impulsive)**.

**Component 17**
Original HCP scores: Gambling Task Reward Percentage Larger (-0.53), Gambling Task Reward Median Reaction Time Larger (0.13), Gambling Task Reward Percentage Smaller (0.53), Gambling Task Reward Median Reaction Time Smaller (0.09), Gambling Task Punish Percentage Larger (0.48), Gambling Task Punish Median Reaction Time Larger (0.12), Gambling Task Punish Percentage Smaller (-0.48), Gambling Task Punish Median Reaction Time Smaller (0.16).

This component is larger for the subjects who gamble less when they will win, and gamble more when they will lose; thus, we named it **Gambling (bad)**.

**Component 18**
Original HCP scores: Endurance Unadjusted (1.0) and Endurance Age-Adjusted (1.0).

This component is larger when subjects are fitter; we named it **Endurance**.

**Component 19**
Original HCP scores: Emotion Task Correct Responses (0.76) and Emotion Task Correct Responses Time (-0.74).

This component is larger in subjects who recognize emotional faces better and faster; we named it **Emotion (good)**.

**Component 20**
Original HCP scores: Emotion Task Correct Angry Responses (0.6), Emotion Task Correct Fear Responses (0.59), Emotion Task Correct Happy Responses (0.06), Emotion Task Correct No Emotion Responses (-0.38) and Emotion Task Correct Sadness Responses (0.59).

This component is greater when subjects better recognize negative emotions, and fail more at recognizing neutral faces. We thus named it **Emotion (negative bias)**.

**Component 21**
This component is only composed of one score: Noise Component. We thus named it **Noise**.

**Component 22**
Original HCP scores: Odor Identification Unadjusted (0.95) and Odor Identification Age-Adjusted (0.95).

This component is larger when subjects better recognize odors; we named it **Odor**.

**Component 23**
Original HCP scores: Pain Intensity Raw Score (0.95) and Pain Intensity T Score (0.94).

This component is larger when subjects feel pain to a greater extent; we named it **Pain**.

**Component 24**
Original HCP scores: Taste Intensity Unadjusted (1.0) and Taste Intensity Age-Adjusted (1.0).

This component is larger when subjects show larger taste intensity; we named it **Taste**.

**Component 25**
Original HCP scores: Mars Contrast Sensitivity Score (0.72), Errors on Mars (-0.41) and Mars Final Contrast Sensitivity Score (0.99).

This component is larger in subjects who exhibit greater sensitivity to visual contrast; we named it **Contrast sensitivity**.

**Component 26**
Original HCP scores: NEO-FFI Agreeableness (-0.66), NEO-FFI Openness to Experience (0.04), NEO-FFI Conscientiousness (-0.71), NEO-FFI Neuroticism (0.73) and NEO-FFI Extraversion (-0.71).

Larger component values indicate subjects who are introverted, neurotic, less conscientious and less agreeable. As such, it bears a negative and an introverted meaning. We called the component **NEOFAC (negative/introverted)**

**Component 27**
Original HCP scores: NEO-FFI Agreeableness (0.17), NEO-FFI Openness to Experience (0.97), NEO-FFI Conscientiousness (-0.2), NEO-FFI Neuroticism (-0.05) and NEO-FFI Extraversion (0.13).

This component showed larger values in subjects more open to experience. We called it **NEOFAC (Openness to experience)**.

**Component 28**
Original HCP scores: ASR Anxiety Raw Score (0.94), ASR Anxiety Percentile (0.88), DSM Anxiety Raw Score (0.95), DSM Anxiety T Score (0.89).

This component was larger in more anxious subjects; we thus called it **Anxiety**.

**Component 29**
Original HCP scores: ASR Withdrawal Symptoms Raw Score (1.0), ASR Withdrawal Symptoms Percentile (0.94).

This component was larger in subjects with a greater extent of withdrawal symptoms; we thus called it **Withdrawal**.

**Component 30**
Original HCP scores: ASR Somatic Problems Raw Score (0.98), ASR Somatic Problems T Score (0.95), DSM Somatic Problems Raw Score (0.95), DSM Somatic Problems T Score (0.95).

This component was larger in subjects with a greater extent of somatic problems; we thus called it **Somatic problems**.

**Component 31**
Original HCP scores: ASR Thought Problems Raw Score (1.0), ASR Thought Problems T Score (0.96).

This component was larger in subjects with a greater extent of thought problems; we thus called it **Thought problems**.

**Component 32**
Original HCP scores: ASR Attentional Problems Raw Score (1.0), ASR Attentional Problems T Score (0.97).

This component was larger in subjects with a greater extent of attentional problems; we thus called it **Attentional problems**.

**Component 33**
Original HCP scores: ASR Aggressiveness Raw Score (1.0), ASR Aggressiveness T Score (0.94).

This component was larger in more aggressive subjects; we thus called it **Aggressiveness**.

**Component 34**
Original HCP scores: ASR Rule Breaking Raw Score (1.0), ASR Rule Braking T Score (0.99).

This component was larger in the subjects showing a greater tendency to break rules; we thus called it **Rule breaking**.

**Component 35**
Original HCP scores: ASR Intrusive Raw Score (1.0), ASR Intrusive T Score (0.97).

This component was larger in more intrusive subjects; we thus called it **Intrusive**.

**Component 36**
Original HCP scores: ASR Internalizing Raw Score (1.0), ASR Internalizing T Score (1.0).

This component was larger in the subjects internalizing more; we thus called it **Internalizing**.

**Component 37**
Original HCP scores: ASR Externalizing Raw Score (1.0), ASR Externalizing T Score (1.0).

This component was larger in the subjects externalizing more; we thus called it **Externalizing**.

**Component 38**
Original HCP scores: ASR Avoidance Raw Score (1.0), ASR Avoidance T Score (0.99).

This component was larger in the subjects showing more avoidance behaviors; we thus called it **Avoidance**.

**Component 39**
Original HCP scores: ASR ADHD Raw Score (1.0), ASR ADHD T Score (0.99).

This component was larger in the subjects with more ADHD traits; we thus called it **ADHD**.

**Component 40**
This component is only composed of one score: ASR Inattention. We thus named it **Inattention**.

**Component 41**
This component is only composed of one score: ASR Hyper-responsiveness. We thus named it **Hyper-responsiveness**.

**Component 42**
Original HCP scores: ASR Antisocial Raw Score (1.0), ASR Antisocial T Score (0.99).

This component was larger in the subjects showing more antisocial behaviors; we thus called it **Antisocial**.

**Component 43**

Original HCP scores: Total Drinks in Last Seven Days (0.81), Number Days Drank in Last Seven Days (0.73), Number of DSM4 Alcohol Dependence Criteria Endorsed (0.52), Number of DSM4 Alcohol Abuse Criteria Endorsed (0.53), Drinks per Drinking Day in Last 12 Months (0.63), Frequency of Alcohol Use in Last 12 Months (-0.79) and Frequency Drunk in Last 12 Months (-0.77).

The last two scores take lower values for greater frequencies; thus, a larger component value indicates more severe alcohol-related problems, and we named the component **Alcohol**.

**Component 44**

Original HCP scores: Total Times Smoked in Last Seven Days (0.68), Times Smoked Today (0.54), Number of Days Smoked in Last Seven Days (0.67), SSAGA Smoking History (0.97).

Greater component values highlight the subjects consuming more tobacco. We named the component **Tobacco**.

**Component 45**

Original HCP scores: Times Used Illicits (0.72), Times Used Cocaine (0.47), Times Used Hallucinogens (0.51), Times Used Opiates (0.47), Times Used Sedatives (0.46), Times Used Stimulants (0.52), Times Used Marijuana (0.92).

Greater component values indicate the subjects who consumed a larger amount of drugs; we thus called the component **Drugs**.

**Robustness of the results to parameter changes**

We verified that modifications of the parameters used in our main analyses would not substantially affect our results. We considered the followings:

1. The threshold (in mm) above which a time point is censored. We compared our main choice of 0.3 mm to values of 0.2 mm (case 1), 0.5 mm (case 2) and 1 mm (case 3).
2. The number of frames to excise around corrupted time points. We compared the case of solely removing the corrupted frames, as done in our main results, to the additional removal of one more frame at time *t+1* (case 4).
3. The number of nearest neighbors used for graph construction; we contrasted our original value of 10 to alternative values of 5 (case 5) and 20 (case 6).
4. The number of time bins into which to subdivide a resting-state session; in comparison with our original choice of 6 bins, we probed values ranging from 4 (case 7) to 8 (case 10).

We assessed robustness of the clustering outcomes by computing the purity measure (Yang et al. 2012), which takes a value of 1 for perfect concordance of classification, and of 0 if no data point is clustered similarly.

For each set of saliences, we computed (1) the absolute valued Spearman's correlation between the reference and output saliences, and (2) their absolute valued dot product. The absolute value enables to account for sign-flipped salience vectors across computations.

|  | Purity | Spearman's correlation | | | | Dot product | | |
| --- | --- | --- | --- | --- | --- | --- | --- | --- |
|  |  |  | $C_1$ | $C_2$ | $C_3$ | $C_1$ | $C_2$ | $C_3$ |
| Case 1 | 0.96 | $S_B$ | 1.00 | 0.99 | 0.99 | 1.00 | 1.00 | 0.99 |
|  |  | $S_M$ | 0.98 | 0.96 | 0.94 | 1.00 | 0.99 | 0.99 |
| Case 2 | 0.96 | $S_B$ | 1.00 | 1.00 | 1.00 | 1.00 | 1.00 | 1.00 |
|  |  | $S_M$ | 0.96 | 0.95 | 0.96 | 1.00 | 1.00 | 1.00 |
| Case 3 | 0.96 | $S_B$ | 1.00 | 1.00 | 0.99 | 1.00 | 1.00 | 0.99 |
|  |  | $S_M$ | 0.95 | 0.93 | 0.92 | 0.99 | 0.99 | 0.99 |
| Case 4 | 0.96 | $S_B$ | 1.00 | 0.99 | 1.00 | 1.00 | 1.00 | 1.00 |
|  |  | $S_M$ | 0.95 | 0.95 | 0.95 | 0.99 | 0.99 | 1.00 |
| Case 5 | 1.00 | $S_B$ | - | - | - | - | - | - |
|  |  | $S_M$ | - | - | - | - | - | - |
| Case 6 | 1.00 | $S_B$ | - | - | - | - | - | - |
|  |  | $S_M$ | - | - | - | - | - | - |
| Case 7 | 0.88 | $S_B$ | 0.99 | 0.99 | 0.98 | 0.99 | 0.98 | 0.97 |
|  |  | $S_M$ | - | - | - | - | - | - |
| Case 8 | 0.86 | $S_B$ | 0.99 | 0.99 | 0.98 | 1.00 | 0.99 | 0.99 |
|  |  | $S_M$ | - | - | - | - | - | - |
| Case 9 | 0.94 | $S_B$ | 0.99 | 0.99 | 0.97 | 1.00 | 0.99 | 0.98 |
|  |  | $S_M$ | - | - | - | - | - | - |
| Case 10 | 0.91 | $S_B$ | 0.99 | 0.97 | 0.95 | 1.00 | 0.98 | 0.95 |
|  |  | $S_M$ | - | - | - | - | - | - |

**Supplementary Table 3:** Robustness of the results to parameter changes. Cases 5 and 6 involved changes of the number of nearest neighbors, which does not alter the generation of PLS components; for this reason, robustness quantification is not provided for this case. Similarly, spatiotemporal features generated with a different number of time bins (cases 7 to 10) do not enable to compare motion saliences. $S_B$, behavioral saliences; $S_M$, motion saliences.

As can be seen in the above table, our results were robust to all the investigated parameter changes.

|              | **Group 1** (70) | **Group 2** (51) | **Group 3** (67) | **Group 4** (36) |
|--------------|------------------|------------------|------------------|------------------|
| Age          | 29.79 (2.55)     | 29.96 (3.1)      | 30.34 (2.91)     | 29.58 (2.55)     |
| Gender (M/F) | 35/35            | 9/42             | 24/43            | 10/26            |
| FD           | 0.13 (0.03)      | 0.14 (0.03)      | 0.25 (0.09)      | 0.15 (0.03)      |

**Supplementary Table 4**: For the four extracted spatiotemporal mover subgroups, age, gender and framewise displacement (FD) values. Standard deviations are provided in brackets, and numbers in brackets beside group legends highlight the number of subjects in each group.